\documentclass[12pt,a4]{article}
\usepackage{setspace} 
\usepackage[ansinew]{inputenc} 
\usepackage{amsthm}

\usepackage[T1]{fontenc} 
\usepackage{lmodern,microtype} 

\usepackage{amsmath,amsthm,amssymb,amsfonts} 
\usepackage{mathrsfs,ushort} 

\usepackage{titlesec,titling} 
\usepackage[nohead]{geometry} 
\usepackage{setspace} 
\usepackage{enumitem,booktabs} 

\usepackage[comma]{natbib} 

\usepackage{hyperref}
\usepackage{bm}
\usepackage{mathtools}
\usepackage{dsfont}
\usepackage{comment}
\usepackage{commath}
\usepackage{colonequals}
\usepackage{mathtools}
\usepackage{cases}
\usepackage{threeparttable}
\usepackage{mathtools}
\usepackage{graphicx}
\usepackage{empheq}
\usepackage{lmodern}
\usepackage{nccmath}
\usepackage{amsmath}

\RequirePackage{natbib}

\newcommand\Item[1][]{%
	\ifx\relax#1\relax  \item \else \item[#1] \fi
	\abovedisplayskip=0pt\abovedisplayshortskip=0pt~\vspace*{-\baselineskip}}

\title{\bfseries\Large A Novel Approach to Predictive Accuracy Testing in Nested Environments \vspace*{-1ex}}

\author{Jean-Yves Pitarakis\thanks{I wish to thank the Editor, Co-Editor and three anonymous referees for the quality of the reports I have received and their in-depth review of an earlier version of this paper. I also wish to thank the ESRC for its financial support via grant ES/W000989/1. Any errors are my own responsibility.   
		Address for Correspondence: Jean-Yves Pitarakis, Department of Economics, University of Southampton, Southampton SO17 1BJ, United-Kingdom. Matlab codes and replication files are available for download from the author's webpage at \url{https://sites.google.com/view/jpitarakis}
} \\
	University of Southampton\\
	Department of Economics\\
	j.pitarakis@soton.ac.uk}

\titleformat{\section}[block]{\centering\large\bfseries}{\thesection.}{0.5em}{}
\titleformat{\subsection}[block]{\flushleft\bfseries}{\thesubsection.}{0.5em}{}
\titleformat{\subsubsection}[runin]{\normalsize\itshape}{\bfseries\thesubsubsection.}{0.5em}{}[.--\:]

\renewcommand{\thesubsubsection}{\arabic{section}.\arabic{subsection}.\alph{subsubsection}}
\titlespacing{\section}{0ex}{6ex}{3ex}
\titlespacing{\subsection}{0in}{3ex}{1.5ex}
\titlespacing{\subsubsection}{0mm}{2ex}{0.5em}
\renewcommand{\linespread}[1]{\setstretch{1}}

\begin{document}

\maketitle

\newpage
\begin{abstract}
We introduce a new approach for comparing the predictive accuracy of two nested models that bypasses the difficulties caused by the degeneracy of the asymptotic variance of forecast error loss differentials used in the construction of commonly used predictive comparison statistics. Our approach continues to rely on the out of sample MSE loss
differentials between the two competing models, leads to nuisance parameter free Gaussian asymptotics and is shown to remain valid under 
flexible assumptions that can accommodate heteroskedasticity and the presence of mixed predictors (e.g. stationary and local to unit root).  A local power analysis also establishes their ability to detect departures from the null in both stationary and persistent settings. 
Simulations calibrated to common economic and financial applications indicate that our methods have strong power with good size control across commonly encountered sample sizes. \\


\end{abstract}	

\vspace{1cm}

\noindent
{\bf Running Head:} PREDICTIVE ACCURACY IN NESTED ENVIRONMENTS

\newpage	
\section{Introduction}

This paper is concerned with comparing the forecasting performance of two nested models through tests that rely on out of sample mean squared error (MSE) loss differentials. Our proposed approach bypasses the  widely documented complications caused by 
the degenerate asymptotic variances of these differentials that occur in nested environments while also leading to 
nuisance parameter free standard normal asymptotics. Our approach  remains valid under both stationary and persistent predictors thus also greatly expanding its practical relevance in economics and finance.

Since the early work of \cite{dm1995} and \cite{w1996} 
a vast body of theoretical research has been concerned with developing new methods for comparing the out of sample predictive ability of competing models. Such tests typically compare the out of sample forecast errors generated from two models under a variety of loss functions and forecasting schemes (e.g. recursive, rolling or fixed updating of model parameter estimates) with the aim of testing the null hypothesis of 
equal predictive accuracy. Most of the test statistics introduced in this literature are based on estimated out of sample MSE loss differentials associated with the two competing forecast error series and have been shown to be asymptotically normally distributed provided that the models being compared are non-nested and a set of standard regularity conditions hold.

The fundamental difficulties that arise as one moves from a non-nested to a nested environment have also generated a vast and growing literature aiming to operationalise and adapt the above approach to nested models. 
In a nested modelling context a key complication comes from the fact that under the null of interest the population errors of the two models are identical thus leading to sample MSE loss differentials that are identically zero in the limit with null asymptotic variances. These in turn result in test statistics that are not well defined asymptotically
and in the failure of normal approximations for popular test statistics such as the Diebold-Mariano statistic (henceforth referred to as DM). 

Alternative normalisations applied to the MSE loss differentials in nested contexts have subsequently been shown to lead to test statistics with well defined but no longer Gaussian limiting null distributions expressed as functionals of stochastic integrals in Brownian Motions (\cite{cm2001, cm2005}, \cite{m2007}, \cite{ht2015}). With the exception of restrictive frameworks that rule out heteroskedasticity 
or allow only a single additional predictor in the nesting model these distributions typically depend on a variety of model specific parameters that cannot be eliminated via standard 
HAC type corrections, requiring simulation based approaches for their implementation (see \cite[pp. 126-127]{w2006}, and \cite[pp. 10-15]{cm2013}). The asymptotics of these test statistics are further influenced by how the in-sample observations are allowed to grow relative to the out of sample observations and the particular choice of the forecasting scheme used to generate forecasts. 

Rather than relying on these non-standard and non-Gaussian distributions this same literature 
has also proposed to bypass the difficulties underlying nested model comparisons by continuing to 
use normal approximations for adjusted versions of DM type statistics. In \cite{cw2007} for instance the authors 
introduced an adjustment to the spread of the out of sample MSEs of the two competing models and argued that although asymptotic normality cannot be established per se such an approach results in reasonably accurate inferences with acceptable size distortions. The adjustment essentially 
corrects for the fact that under the null hypothesis of equal predictive accuracy the MSE of the larger model is contaminated with estimation noise. This adjusted DM type statistic proposed in \cite{cw2007} has become the norm in economic applications 
involving out of sample forecast comparisons with recent examples found in \cite{mp2009}, \cite{imp2016}, \cite{ew2021} amongst numerous others. 

In this paper we introduce an alternative formulation of the out of sample MSE loss differential between two models that is not subject to the variance degeneracy problem of existing procedures. This subsequently allows us to construct novel test statistics for testing the null hypothesis of equal out of sample population MSEs which are shown to have simple nuisance parameter free normal distributions. 
The main idea underlying our proposed approach is based on the observation that MSE comparisons across two competing models need not be performed within the same out of sample span of available forecast error observations. These can be performed over partially overlapping segments instead, leading to test statistics that  
accumulate MSE spreads over all possible such segments. This new setting can trivially accommodate desirable features such as conditional heteroskedasticity and persistent predictors 
and is also shown to lead to both consistent and locally powerful test statistics. As we discuss further below, our approach can also be adapted to broader contexts where the nestedness of models is an important consideration for inferences such as model selection testing.  

Besides conventional forecasting objectives, nested models are commonly encountered environments when it comes to testing economic hypotheses and validating theories. Notable examples include forecast accuracy comparisons against random walk models in the exchange rate literature spurred by the early work of \cite{mr1983} and more recently reconsidered in \cite{r2005}, \cite{mp2009} amongst others, equity premium predictability issues as recently investigated in \cite{fnb2013}, \cite{aw2017} and numerous others. 
Our key aim here is to propose a way of addressing and resolving a long standing issue that has generated a vast agenda on the formal comparison of such models via their out-of-sample predictive accuracy. The important auxiliary debate on the advantages or disadvantages of using out-of-sample versus in-sample approaches is 
not part of our focus. 
It is also important to emphasise that our interest here is on testing population level predictive ability when forecasts are generated recursively as opposed to finite sample based predictive ability
as considered for instance in \cite{gw2006}. This latter approach is 
able to avoid the complications induced by the nestedness of models being compared by proceeding via a rolling fixed window based forecasting scheme 
so that the issue of 
competing models becoming identical in the limit can be bypassed. 

Throughout this paper we also followed the common practice of referring to statistics based on MSE differentials 
obtained from competing estimated models as Diebold-Mariano type statistics. We must acknowledge however that the specific testing approach initially developed by these authors was not concerned with model specific considerations or specification testing motives as its underlying theory was developed for \emph{given} sequences of forecast errors assumed to satisfy certain regularity conditions (see \cite{d2015}). Nevertheless, the forecasting literature of the past decade has generally amalgamated the notion of  forecast evaluation with the evaluation of models on the basis of their forecasting abilities. 

The paper is organised as follows. Section 2 introduces the nested forecasting environment and establishes the limiting null distributions of two novel test statistics. Section 3 concentrates on their asymptotic power properties, establishing their consistency and ability to detect local departures from the null. Section 4 introduces a simple adjustment to the same statistics shown to further enhance their power properties without affecting their 
null distributions. Section 5 provides a comprehensive finite sample evaluation of our methods based on two DGPs calibrated to commonly encountered applications. Section 6 illustrates the use of our proposed methods 
via an application to exchange rate models. Section 7 overviews our key results and discusses extensions. Proofs are given in the Appendix. Further simulation results are provided in an online supplement. 

\section{Models and Test Statistics: Theory}

\noindent 
We consider the following predictive regressions
\begin{ceqn}
\begin{align}
y_{t+1} & =  {\bm x}_{1t}'{\bm \delta}_{1}+v_{t+1} \label{eq:1} \\
y_{t+1} & =  {\bm x}_{1t}'{\bm \beta}_{1} \label{eq:2}
+{\bm x}_{2t}'{\bm \beta}_{2}+u_{t+1} 
\end{align}
\end{ceqn}
where the ${\bm x}_{it}$'s are the $(p_{i}\times 1)$ vectors of predictors,  ${\bm \delta}_{1}$ and ${\bm \beta}_{i}$ the $(p_{1}\times 1)$ and $(p_{i}\times 1)$ parameter vectors and $v_{t}$ and $u_{t}$ the random disturbance terms. We let ${\bm x}_{t}=({\bm x}_{1t}',{\bm x}_{2t}')'$ and ${\bm \beta}=({\bm \beta}_{1}',{\bm \beta}_{2}')'$ and set $p=p_{1}+p_{2}$. Here model (\ref{eq:1}) is nested within the larger model in (\ref{eq:2}) and under ${\bm \beta}_{2}=0$ we have ${\bm \delta}_{1}\equiv {\bm \beta}_{1}$ and $v_{t+1}\equiv u_{t+1}$. 
The formulation of the above two nested models is standard and parallels closely the most commonly encountered setting 
considered in the predictive accuracy testing literature as for instance in \cite{ht2015}.

One step ahead forecasts of $y_{t+1}$ from (\ref{eq:1}) and (\ref{eq:2}) are generated recursively as $\hat{y}_{1,t+1|t}={\bm x}_{1t}'\hat{\bm \delta}_{1t}$
and $\hat{y}_{2,t+1|t}={\bm x}_{t}'\hat{\bm \beta}_{t}$ for $t=k_{0},\ldots,T-1$
where
$\hat{\bm \delta}_{1t}=(\sum_{j=1}^{t}{\bm x}_{1j-1}{\bm x}_{1j-1}')^{-1}\sum_{j=1}^{t}{\bm x}_{1j-1}'y_{j}$, $\hat{\bm \beta}_{t}=(\sum_{j=1}^{t}{\bm x}_{j-1}{\bm x}_{j-1}')^{-1}\sum_{j=1}^{t}{\bm x}_{j-1}'y_{j}$ 
and the resulting pseudo out of sample forecast errors are then obtained as
$\hat{e}_{1,t+1}=y_{t+1}-{\bm x}_{1t}'\hat{\bm \delta}_{1t}$
and
$\hat{e}_{2,t+1}=y_{t+1}-{\bm x}_{t}'\hat{\bm \beta}_{t}$. Here $k_{0}$ is the sample location used to initiate the first recursive forecasts that lead to the first out of sample forecast errors $\hat{e}_{1,k_{0}+1}$ and $\hat{e}_{2,k_{0}+1}$
and subsequently resulting in $(T-k_{0})$ out of sample forecast error observations. Throughout this paper we take $k_{0}$ to be a given fraction of the sample size, setting $k_{0}=[T \pi_{0}]$ for some $\pi_{0} \in (0,1)$.

Following the early work of \cite{dm1995}, \cite{w1996} and others, a common approach for comparing the predictive accuracy of the two models under MSE loss involves testing the null hypothesis
\begin{ceqn}
\begin{align}
H_{0} & \colon  E[y_{t+1}-\hat{y}_{1,t+1}(\bm \delta_{1})]^{2}=E[y_{t+1}-\hat{y}_{2,t+1}(\bm \beta)]^{2} \label{eq:3}
\end{align}
\end{ceqn}
using a test statistic based on suitably normalised versions of the sample average MSE loss differentials 
\begin{ceqn}
\begin{align}
\overline{D}_{T} & =  \frac{1}{T-k_{0}} \left(\sum_{t=k_{0}}^{T-1}\hat{e}_{1,t+1}^{2}-\sum_{t=k_{0}}^{T-1} \hat{e}_{2,t+1}^{2}\right). \label{eq:4}
\end{align}
\end{ceqn}

Within a non-nested setting and a strictly stationary and ergodic environment \cite{dm1995} and  \cite{w1996} established a standard normal limit theory for this class of test statistics 
(e.g. $\sqrt{T-k_{0}} \ \overline{D}_{T}/\hat{\sigma}_{{\overline{D}}_{T}}$ with $\hat{\sigma}^{2}_{{\overline{D}}_{T}}$ denoting some suitable long run variance estimator) leading to their systematic use in applied work and a voluminous literature on their refinements. Within a nested context  
where $u_{t+1}\equiv v_{t+1}$ however it is straightforward to observe that
$\sqrt{T-k_{0}} \ \overline{D}_{T}\stackrel{p}\rightarrow 0$ and 
$\hat{\sigma}_{{\overline{D}}_{T}} \stackrel{p}\rightarrow 0$ invalidating the limiting standard normal approximation and the use of these test statistics for inference purposes. This degeneracy problem is not solely confined to 
the Diebold-Mariano type statistics but universally affects all existing methods
that compare forecast errors (or models) in nested settings with recursively generated forecasts. 

These observations have
 led to a vast body of research on out of sample predictive accuracy testing in nested models due to their importance in empirical applications in areas such as asset pricing and the modelling of expected returns in particular. 
 For their validity, inferences in nested contexts such as (\ref{eq:1}) and (\ref{eq:2}) must 
 rely on the observation that under $H_{0}$ it is $(T-k_{0})\overline{D}_{T}$ rather than 
 $\sqrt{T-k_{0}} \ \overline{D}_{T}$ that turns out to have a non-degenerate limit 
 which could be used for developing suitable inferences
 (see \cite{cm2001, cm2005}, \cite{m2007}, \cite{ht2015}). 
  This however is also problematic due to the non-standard and non-pivotal nature of 
 the resulting asymptotic distributions.
These take the form of functionals of stochastic integrals in Brownian Motions and with the exception of some special cases 
 contain nuisance parameters that are difficult to remove via 
 standard HAC type normalisations. Even under special instances such as conditional homoskedasticity these distributions continue to depend on the number of extra predictors included in the nesting models and the fraction of the sample used to build the first recursive forecasts. Equally importantly these 
 results have been obtained under stationarity and ergodicity assumptions ruling out the important and frequently encountered case of predictors having roots near unity in their autoregressive representations. 
  
Instead of evaluating the two sequences of squared forecast errors
 $\{\hat{e}_{1,t+1}^{2}\}$ and 
 $\{\hat{e}_{2,t+1}^{2}\}$ over the entire {\it and} same interval
 $[k_{0}+1,T]$ as it is done in the formulation of all commonly used 
 test statistics based on $\overline{D}_{T}$
 we here propose to compare the two out of sample MSEs over partially overlapping segments of the $[k_{0}+1,T]$ interval instead. 
 For this purpose 
 we introduce the following generalised MSE spread 
 \begin{ceqn}
 \begin{align}
 	\widetilde{D}_{T} (\ell_{1},\ell_{2}) & =  \dfrac{\sum\limits_{t=k_{0}}^{k_{0}+\ell_{1}-1}\hat{e}_{1,t+1}^{2}}{\ell_{1}}-\dfrac{\sum\limits_{t=k_{0}}^{k_{0}+\ell_{2}-1} \hat{e}_{2,t+1}^{2}}{\ell_{2}} \label{eq:5}
 \end{align}
 \end{ceqn}
where $\ell_{1}$ and $\ell_{2}$ control the range over which the two squared forecast error sequences are evaluated. 
Note that setting $\ell_{1}=\ell_{2}=T-k_{0}$ 
in (\ref{eq:5}) reduces it to $\overline{D}_{T}$ which can be viewed as a special case of $\widetilde{D}_{T}(\ell_{1},\ell_{2})$. In line with the analysis based on (4) we take $\ell_{1}=[(T-k_{0})\lambda_{1}]$ and $\ell_{2}=[(T-k_{0})\lambda_{2}]$ with $\lambda_{1}$ and $\lambda_{2}$ referring to the fraction of the $(T-k_{0})$ squared forecast errors associated with models (\ref{eq:1}) and (\ref{eq:2}) respectively. 

From a theoretical standpoint, 
proceeding with the use of $\widetilde{D}_{T}(\ell_{1},\ell_{2})$ 
instead of $\overline{D}_{T}$ has no bearing on the null hypothesis being 
tested in the sense that 
when $u_{t+1}\equiv v_{t+1}$ the population counterpart of 
$\widetilde{D}_{T}(\ell_{1},\ell_{2})$ also equals zero. A key feature of (\ref{eq:5}) that distinguishes it from $\overline{D}_{T}$ however is that the variance of its suitably normalised version will no longer be degenerate provided that $\ell_{1}\neq \ell_{2}$ (equivalently $\lambda_{1}\neq \lambda_{2}$). This normalised version of (\ref{eq:5}) which forms the basis of our proposed test statistics is given by $Z_{T}(\ell_{1},\ell_{2})=\sqrt{T-k_{0}} \ \widetilde{D}_{T}(\ell_{1},\ell_{2})$,
\begin{ceqn}
\begin{align}
	{Z}_{T}(\ell_{1},\ell_{2}) & = \frac{T-k_{0}}{\ell_{1}}\left[ \frac{\sum\limits_{t=k_{0}}^{k_{0}+\ell_{1}-1}\hat{e}_{1,t+1}^{2}}{\sqrt{T-k_{0}}}-\frac{\ell_{1}}{\ell_{2}}
	\frac{\sum\limits_{t=k_{0}}^{k_{0}+\ell_{2}-1}\hat{e}_{2,t+1}^{2}}{\sqrt{T-k_{0}}}\right]. \label{eq:6}
\end{align}
\end{ceqn}
Note that (\ref{eq:6}) is simply the normalised difference in the means of the two sample MSEs evaluated over the two relevant segments of the effective sample size. \\

\noindent REMARK 1: A key point to observe here is that 
the variance of (\ref{eq:6}) is well-defined and no longer collapses to zero in the limit 
provided that $\lambda_{1}$ and $\lambda_{2}$ are bounded away from zero and bounded away from each other. 
To illustrate and motivate this point heuristically let us replace both $\hat{e}_{1t+1}^{2}$ and $\hat{e}_{2t+1}^{2}$ in (\ref{eq:6}) with $(u_{t+1}^{2}-\sigma^{2}_{u})$ for $\sigma^{2}_{u}\equiv E[u_{t}^{2}]$.  Taking the $u_{t}'s$ to be IID(0,1) with $E[u_{t+1}^{4}]<\infty$ it follows that
\begin{ceqn}
	\begin{align}
	V[{Z}_{T}(\ell_{1},\ell_{2})] & \rightarrow  V[u_{t+1}^{2}] \ \dfrac{|\lambda_{2}-\lambda_{1}|}{\lambda_{1}\lambda_{2}} \label{eq:7}
	\end{align}
\end{ceqn}
suggesting that a test statistic based on $\widetilde{D}_{T}(\ell_{1},\ell_{2})$ will  not have a degenerate distribution as it was the case with the use 
of $\overline{D}_{T}$ in nested contexts. We may also wish to point out that having $\lambda_{1}$ and $\lambda_{2}$ bounded away from zero is merely a technical requirement in the asymptotics that follow as in practice these parameters will naturally be set at or near their maximum boundary of one. 

The quantity in (\ref{eq:6}) forms the building block of our proposed test statistics for testing the null hypothesis in (\ref{eq:3}) against one sided right tail based alternatives as it is the norm in this literature. We consider two types of test statistics that operationalise (\ref{eq:6}). Our choice is guided by the simplicity of the ensuing asymptotics and their intuitive interpretation 
while recognising that alternative constructions/normalisations of $\widetilde{D}_{T}(\ell_{1},\ell_{2})$ may also be 
considered. 

The first test statistic that we consider is denoted $Z_{T}^{0}(\lambda_{1}^{0},\lambda_{2}^{0})$ and is based on implementing inferences for given magnitudes $\ell_{1}^{0}=[(T-k_{0})\lambda_{1}^{0}]$ and $\ell_{2}^{0}=[(T-k_{0})\lambda_{2}^{0}]$. We write 
\begin{ceqn}
\begin{align}
Z_{T}^{0}(\lambda_{1}^{0},\lambda_{2}^{0}) & =  \frac{1}{\hat{\sigma}} \  Z_{T}([(T-k_{0})\lambda_{1}^{0}],[(T-k_{0})\lambda_{2}^{0}]) \label{eq:8}
\end{align}
\end{ceqn}
with $\hat{\sigma}^{2}$ denoting a consistent estimator of $V[u_{t+1}^{2}]$. 

Our second test statistic is based on averaging (\ref{eq:6}) across the $\ell_{j}$'s. 
The averaging can be implemented over $\ell_{1}\in [1,T-k_{0}]$ for a given $\ell_{2}^{0}$ (e.g., $\ell_{2}^{0}=T-k_{0}$) so that the MSE of the smaller model accumulates progressively as $\ell_{1}$ increases. More generally, this averaging 
can be performed over any desired and feasible range of $\ell_{1}$. To allow such level of generality we introduce the fractional parameter $\tau_{0}$ 
and write
\begin{ceqn}
\begin{align}
	\overline{Z}_{T}(\tau_{0}; \lambda_{2}^{0}) & =  \frac{1}{\hat{\sigma}}\frac{1}{[(T-k_{0})(1-\tau_{0})]}\sum_{\ell_{1}=[(T-k_{0})\tau_{0}]+1}^{T-k_{0}} Z_{T}(\ell_{1},[(T-k_{0})\lambda_{2}^{0}]) \label{eq:9}
\end{align}
\end{ceqn}
where the choice of $\tau_{0}$ determines the user-chosen range of $\ell_{1}$ over which 
the average of $Z_{T}(\ell_{1}, \ell_{2}^{0})$ is taken (given $\ell_{2}^{0}=[(T-k_{0})\lambda_{2}^{0}]$). 
Rather than imposing a fixed and given $\lambda_{1}^{0}$
as in (\ref{eq:8}) this average based statistic essentially considers a range of such magnitudes
and subsequently aggregates outcomes via averaging. Given the role played by $\ell_{1}$ and $\ell_{2}$ in our inferences we can expect that choosing the averaging range in a way that excludes low magnitudes of $\ell_{1}$ (so that the estimated MSEs associated with model 1 remain sufficiently accurate) will result in more reliable inferences.   
  The issue of how best to select these user-inputs is postponed until Section 3 where we provide precise guidelines informed by a theoretical local power analysis. One motivation behind this average based statistic when compared with (\ref{eq:8}) is that one can remain partly more agnostic about the specific magnitude to use for one of the two required user-inputs in $Z_{T}([(T-k_{0})\lambda_{1}^{0}],[(T-k_{0})\lambda_{2}^{0}])$ while setting the other one (e.g., $\lambda_{2}^{0}$) at or near its maximum boundary of one. Although our context is different, this is also reminiscent of the various approaches used in the structural break testing literature when one does not wish to take a stance on the location of a potential change-point. More importantly, and borrowing from the same literature, we may also conjecture that the averaging process may result in tests with more favorable size-power trade-offs.  

At this stage it is also important to point out that there are numerous alternative possibilities for designing 
test statistics in the spirit of (\ref{eq:8}) and (\ref{eq:9}) (e.g., double averaging across $\ell_{1}$ and $\ell_{2}$, alternative functional forms etc.). An interesting avenue for future research could be the design of a class of test statistics 
based on $Z_{T}(\ell_{1},\ell_{2})$ and having desirable optimality properties as it has been attempted in the structural break literature. 

Although both (\ref{eq:8}) and (\ref{eq:9}) allow for a broad range of theoretically feasible magnitudes for $(\lambda_{1}^{0},\lambda_{2}^{0})$ in $Z^{0}_{T}(\lambda_{1}^{0},\lambda_{2}^{0})$ and $(\tau_{0},\lambda_{2}^{0})$ in $\overline{Z}_{T}(\tau_{0};\lambda_{2}^{0})$ one naturally expects that choosing $(\lambda_{1}^{0},\lambda_{2}^{0})$ and $(\tau_{0},\lambda_{2}^{0})$ to lie in the vicinity of unity would capture the greatest amount of information from the two competing models. 
As we show further below such a choice does indeed lead to remarkably powerful tests with excellent size control. Given a sequence of forecast errors available to the investigator the 
practical implementation of either (\ref{eq:8}) or (\ref{eq:9}) is also as straightforward as calculating standard DM type test statistics. 

To establish the limiting properties of our test statistics under the null hypothesis in (\ref{eq:3}) we introduce a set of high level assumptions ensuring a flexible environment that encompasses the vast majority of settings considered in the literature while also allowing for a richer temporal structure.  
As we wish to highlight the generality and usefulness of our methods based on the use of (\ref{eq:5})-(\ref{eq:6}) we abstain from primitive conditions that may unnecessarily suggest a restrictive scope for their use. More importantly our use of high level assumptions is motivated by the fact that our proposed methods can be immediately seen to be robust to a very rich dynamic structure of predictors 
including highly persistent processes, strictly stationary and ergodic processes, long memory processes etc.\\
 
\noindent
ASSUMPTION A: \\
{\em

\noindent
(i) $\displaystyle \sup_{\lambda \in (0,1]} \left|
		\frac{\sum\limits_{t=k_{0}}^{k_{0}-1+[(T-k_{0})\lambda]}\hat{e}_{j,t+1}^{2}}{\sqrt{T-k_{0}}}-
		\frac{\sum\limits_{t=k_{0}}^{k_{0}-1+[(T-k_{0})\lambda ]}u_{t+1}^{2}}{\sqrt{T-k_{0}}}
		\right| \stackrel{H_{0}}=o_{p}(1)$ for $j=1,2$.\\

\noindent		
(ii) The sequence of demeaned squared errors $\eta_{t}=u_{t+1}^{2}-\sigma^{2}_{u}$ has autocovariances 
		$\gamma_{j}^{\eta}$ that satisfy $\sum_{j=0}^{\infty}|\gamma_{j}^{\eta}|<\infty$ and fulfills  
		a functional central limit theorem, that is $T^{-\frac{1}{2}}\sum_{t=1}^{[Ts]} (u_{t+1}^{2}-\sigma^{2}_{u})\stackrel{\cal{D}}\rightarrow  \sigma W_{\eta}(s)$ on $D_{\mathbb{R}}([0,1])$ the space of cadlag functions on $[0,1]$ with $W_{\eta}(.)$ denoting a standard Brownian Motion and $\sigma^{2}=\gamma_{0}^{\eta}+2 \sum_{j=1}^{\infty}\gamma_{j}^{\eta}$>0. \\

\noindent 
(iii) A consistent estimator $\hat{\sigma}^{2}$ of $\sigma^{2}$ exists, that is $\hat{\sigma}^{2} \stackrel{p}\rightarrow \sigma^{2} \in (0,\infty)$.\\
}

\noindent
We note that condition A(i) holds for both $j=1$ and $j=2$ highlighting the fact that we operate within a nested environment with (\ref{eq:1}) being the true model. 
A1(i) is trivially satisfied under a very broad range of settings used to obtain the large sample properties of DM type statistics in nested models. An important feature to also highlight here is the fact that A(i) does not restrict the persistence properties of the predictors which could be highly persistent in the sense of following local to unit root processes for instance. This greatly expands and enriches the environment in
which predictive accuracy inferences have commonly been introduced. 
The robustness of A(i) to the persistence properties of the predictors is an important and useful feature of the squared forecast errors as opposed to their level for which a result such as A(i) would not hold.
For more 
primitive conditions illustrating specialised environments under which A(i) holds see
\cite{dp2008a} and \cite{ht2015} for strictly stationary and ergodic/mixing settings and \cite{brn2019} for environments where A(i) is shown to hold under both stationary and unit-root or near unit-root regressors. 

Assumption A(ii) requires the centered squared errors driving (\ref{eq:1})-(\ref{eq:2}) to satisfy a functional central limit theorem with $\sigma^{2}$ referring to their long run variance. The absolute summability of the autocovariances of $\eta_{t}$ ensures that $\sigma^{2}$ the limit of $V[\sum_{t=k_{0}}^{T-1}\eta_{t+1}/\sqrt{T-k_{0}}]$ exists. Examples of processes 
which satisfy Assumption A(ii) include a broad range of conditionally heteroskedastic ARCH/GARCH processes under suitable existence of moments restrictions. For a detailed set of primitive assumptions ensuring that the stated FCLT holds see \cite[Theorem 5.1]{gkl2000}, \cite[Example 2.2 and Theorem 2.1]{gkl2001}, \cite{bhh2008}, 
\cite{l2009} and references therein. 

Assumption A(iii) requires that the long run variance of 
$\eta_{t}$ be estimated consistently. Such an estimator could be trivially 
constructed using least squares residuals from either the null or alternative models. 
Under conditional homoskedasticity an obvious candidate would be 
$\hat{\sigma}^{2}_{hom}=\sum_{t=k_{0}}^{T-1}\hat{\eta}_{t+1}^{2}/(T-k_{0})$ while under dependent errors  (e.g. if the $u_{t}'s$ follow a GARCH type process) a Newey-West type formulation as in \cite{dp2008b} would be suitable and ensure that A(iii) holds. \\

\noindent
REMARK 2: As pointed out in Remark 1, our asymptotic theory for $Z_{T}^{0}(\lambda_{1}^{0},\lambda_{2}^{0})$ in (\ref{eq:8}) imposes $\lambda_{1}^{0}$ and $\lambda_{2}^{0}$ to be bounded away from zero and to be bounded away from each other, say $0<\underline{\lambda}\leq \lambda_{i}^{0}\leq 1$ for $i=1,2$ and $|\lambda_{1}^{0}-\lambda_{2}^{0}| \geq \epsilon$ for some positive fraction $\epsilon$. In what follows we refer to such a set from which these two user-inputs can be selected as $\Lambda^{0}$. The implementation of the average based statistic $\overline{Z}_{T}(\tau_{0};\lambda_{2}^{0})$ in (\ref{eq:9}) requires setting $\lambda_{2}^{0}$ as above and averaging 
$Z_{T}(\ell_{1},[(T-k_{0})\lambda_{2}^{0}])$ across $\ell_{1}=[(T-k_{0})\tau_{0}]+1,\ldots,(T-k_{0})$ for some $\tau_{0}$ bounded away from zero and one. We refer to this set as $\overline{\Lambda}^{0}$. \vspace{0.2cm}

The following two propositions summarise the large sample behavior of our two test statistics under the null hypothesis stated in (\ref{eq:3}). 
\vspace{0.2cm}

\noindent
{\bf PROPOSITION 1}. {\em Under Assumptions A(i)-(iii), the null hypothesis in (\ref{eq:3}), 
and for given $(\lambda_{1}^{0},\lambda_{2}^{0})\in \Lambda^{0}$, we have as $T \rightarrow \infty$
\begin{ceqn}
\begin{align}
	{Z}_{T}^{0}(\lambda_{1}^{0},\lambda_{2}^{0})  & \stackrel{\cal{D}}\rightarrow \mathcal{N} (0,v^{0}(\lambda_{1}^{0},\lambda_{2}^{0})) \label{eq:10}
\end{align}
\end{ceqn}
where
\begin{ceqn}
\begin{align}
v^{0}(\lambda_{1}^{0},\lambda_{2}^{0}) & =  \dfrac{|\lambda_{1}^{0}-\lambda_{2}^{0}|}{\lambda_{1}^{0}\lambda_{2}^{0}}. \label{eq:11}
\end{align}
\end{ceqn}
}

\noindent
{\bf PROPOSITION 2}. {\em Under Assumptions A(i)-(iii), the null hypothesis in (\ref{eq:3}), and for given $(\tau_{0},\lambda_{2}^{0}) \in \overline{\Lambda}^{0}$,  we have as $T \rightarrow \infty$ 
\begin{ceqn}
\begin{align}
	\overline{Z}_{T}(\tau_{0};\lambda_{2}^{0}) & \stackrel{\cal{D}}\rightarrow  \mathcal{N}(0,\bar{v}(\tau_{0};\lambda_{2}^{0})) \label{eq:12}
\end{align}
\end{ceqn}
where 
	{\small
	\begin{numcases}{\hspace*{-1.88em} \bar{v}(\tau_{0};{\lambda_{2}}^{0})=}
	\dfrac{(1-\tau_{0})^{2}+2\lambda_{2}^{0}(1-\tau_{0}+\ln \tau_{0})}{\lambda_{2}^{0}(1-\tau_{0})^{2}} & 
	$\lambda_{2}^{0}\leq \tau_{0}$ \label{eq:13} \hspace*{-1em} \\
	\dfrac{1-\tau_{0}^{2}+2\lambda_{2}^{0}((1-\tau_{0})\ln \lambda_{2}^{0}+\tau_{0}\ln \tau_{0})}{\lambda_{2}^{0}(1-\tau_{0})^{2}} & 
	$\lambda_{2}^{0}>\tau_{0}.$ \label{eq:14} \hspace*{-1em}
	\end{numcases}
}
}

The variance components of the distributional outcomes in (\ref{eq:10}) and (\ref{eq:12}) are of course known to the investigator so that both $Z_{T}^{0}(\lambda_{1}^{0},\lambda_{2}^{0})$ and  $\overline{Z}_{T}(\tau_{0};\lambda_{2}^{0})$ can be 
trivially standardised as 
\begin{ceqn}
\begin{align}
{\cal S}^{0}_{T}(\lambda_{1}^{0},\lambda_{2}^{0}) & \equiv  \dfrac{Z_{T}^{0}(\lambda_{1}^{0},\lambda_{2}^{0})}{\sqrt{v^{0}(\lambda_{1}^{0},\lambda_{2}^{0})}} \label{eq:15} 
\end{align}
\end{ceqn}
and
\begin{ceqn}
\begin{align}
\overline{{\cal S}}_{T}(\tau_{0};\lambda_{2}^{0}) & \equiv 
\dfrac{\overline{Z}_{T}(\tau_{0};\lambda_{2}^{0})}{\sqrt{\bar{v}(\tau_{0};\lambda_{2}^{0})}} \label{eq:16}
\end{align}
\end{ceqn}
to proceed with standard normal inferences for testing $H_{0}$. 

The results in (\ref{eq:10})-(\ref{eq:14}) highlight the simplicity and practicality of our proposed inferences while at the same time offering a solution to an important problem that has not been satisfactorily resolved in this literature. 
The test statistics in (\ref{eq:15}) and (\ref{eq:16}) allow us to generalise the widely used DM 
style forecast accuracy testing approach to a broad class of empirically relevant models including specifications with highly persistent predictors with or without conditional heteroskedasticity. 

We naturally expect the quality of inferences (e.g. power, size vs power trade-offs) to 
be influenced by the specific choices of ($\lambda_{1}^{0},\lambda_{2}^{0}$) in (\ref{eq:15}) and ($\tau_{0},\lambda_{2}^{0}$) in (\ref{eq:16}). Although the above null asymptotics hold under a very broad range of parameterisations for those user-inputs a formal analysis of their local asymptotic power allows us to 
provide precise and tight guidelines ensuring excellent power properties with good size control.  

\section{Asymptotic Power and Test Parameterisations}

We here deviate from Assumption A(i) in order to evaluate the large sample behavior of ${\cal S}_{T}^{0}(\lambda_{1}^{0},\lambda_{2}^{0})$ and $\overline{{\cal S}}_{T}(\tau_{0};\lambda_{2}^{0})$ when the DGP is given by (\ref{eq:2}). Assumption A(i) continues to hold for $j=2$ but no longer for $j=1$ since model (\ref{eq:1}) is misspecified due to the omitted ${\bm x}_{2,t}$ predictors. As $\hat{e}_{1,t+1}^{2}$ will now be contaminated by those omitted predictors we expect the stochastic properties of the latter (e.g. the variance of the ${\bm x}_{2,t}$'s and their correlation with the ${\bm x}_{1,t}$'s) to influence the power properties of both test statistics. Unlike their null distributions we thus also expect the test statistics to diverge at different rates depending on whether the predictors are stationary or highly persistent. 
Our analysis of the consistency and local power properties of ${\cal S}_{T}^{0}(\lambda_{1}^{0},\lambda_{2}^{0})$ and $\overline{{\cal S}}_{T}(\tau_{0};\lambda_{2}^{0})$ is guided 
by these two distinct scenarios which we consider separately.

\subsection{Consistency and Local Power under Stationarity}

We initially concentrate on the case where the predictors driving both (\ref{eq:1}) and (\ref{eq:2}) are stationary and ergodic. Specifically, we operate under the following set of high level assumptions that mirror closely the most common environments 
considered in the predictive accuracy testing literature. \\

\noindent
ASSUMPTION B1: \\

 {\em 
\noindent
(i) $\displaystyle \sup_{\lambda\in [0,1]}\left\lVert \dfrac{\sum_{t=1}^{[T\lambda]} {\bm x}_{t}{\bm x}_{t}'}{T}-\lambda \ {\bm Q}\right\rVert=o_{p}(1)$
		with ${\bm Q}$ a $p\times p$ nonrandom positive definite matrix, \vspace{0.34cm}

\noindent	
(ii) $\displaystyle \dfrac{\sum_{t=1}^{[T\lambda]} {\bm x}_{t}u_{t+1}}{\sqrt{T}} \stackrel{{\cal D}}\rightarrow {\bm \Omega}^{1/2} \ {\bm W}(\lambda)$ with ${\bm W}(.)$ denoting a p-dimensional standard Brownian Motion and ${\bm \Omega}=E[{\bm x}_{t}{\bm x}_{t}'u_{t+1}^{2}]>0$, \vspace{0.34cm}

\noindent
(iii) Assumptions A(ii)-(iii) hold. \vspace{0.34cm}

\noindent
(iv) The user-inputs in ${\cal S}_{T}(\lambda_{1}^{0},\lambda_{2}^{0})$ and 
$\overline{\cal S}_{T}(\tau_{0};\lambda_{2}^{0})$ are such that 
$(\lambda_{1}^{0},\lambda_{2}^{0})\in \Lambda^{0}$ and $(\tau_{0},\lambda_{2}^{0})\in \overline{\Lambda}^{0}$ respectively. \\

}

\noindent Assumptions B1(i)-(iii) mirror closely the environment of \cite{ht2015} and can be viewed as more primitive conditions ensuring that Assumption A(i) holds. B1(i) requires that the predictors satisfy a uniform law of large numbers and rules out trending or local to unit root predictors while B1(ii) 
ensures that $\{{\bm x}_{t}u_{t+1}\}$ satisfies a multivariate functional central limit theorem. Our main result regarding the asymptotic power properties of the two tests within such a stationary environment is now summarised in Proposition 3 below. \\

\noindent 
{\bf PROPOSITION 3}.  {\it (i) Suppose model (\ref{eq:2}) holds with ${\bm \beta}_{2}\neq 0$ and fixed, then under assumption B1 and as $T \rightarrow \infty$ we have 
	${\cal S}_{T}^{0}(\lambda_{1}^{0},\lambda_{2}^{0}) \stackrel{p}\rightarrow \infty$ and $\overline{{\cal S}}_{T}(\tau_{0};\lambda_{2}^{0}) \stackrel{p}\rightarrow \infty$. (ii)  Suppose model (\ref{eq:2}) holds with ${\bm \beta}_{2}={\bm \gamma}/T^{1/4}$ for ${\bm \gamma\neq 0}$. Under assumption B1, $\lim_{||{\bm \gamma}||\rightarrow \infty}\lim_{T \rightarrow \infty}{\cal S}_{T}^{0}(\lambda_{1}^{0},\lambda_{2}^{0})=\infty$ and $\lim_{||{\bm \gamma}||\rightarrow \infty}\lim_{T \rightarrow \infty}\overline{{\cal S}}_{T}(\tau_{0};\lambda_{2}^{0})=\infty$ in probability.} \\

The above results highlight the consistency of both test statistics as well as their ability to detect local 
departures from the null hypothesis under stationary settings. It is here also important to point out that the local to the null 
parameterisation of ${\bm \beta}_{2}$ based on $T^{1/4}$ rather than the usual $T^{1/2}$ rate commonly encountered in stationary settings is not in any way due to our specific test statistics or assumptions. The same scenario would also occur in a conventional regression based testing 
environment and is due to the fact that we are dealing with inferences about the behavior of {\it squared} errors rather than their level. 

To gain further insights into the specific role played by key factors influencing power it is useful to also present the explicit asymptotic local power functions of the two tests for a given size $\alpha \in (0,1)$. These will in turn be used to provide explicit guidance on selecting suitable parameterisations of our two test statistics (i.e. $(\lambda_{1}^{0},\lambda_{2}^{0})$ in (\ref{eq:15})
and $(\tau_{0};\lambda_{2}^{0})$ in (\ref{eq:16})). In what follows it is useful to also recall that $\pi_{0}$ refers to the given fraction of the sample size used to initiate the recursive computation of forecasts. \\

\noindent
{\bf COROLLARY 1}. {\it  Suppose model (\ref{eq:2}) holds with ${\bm \beta}_{2}={\bm \gamma}/T^{1/4}$ for ${\bm \gamma\neq 0}$. Under assumption B1 and letting $q_{\alpha}$ denote the upper $\alpha$-quantile of the standard normal distribution with CDF $\Phi(.)$, the asymptotic local power functions of the tests based on ${\cal S}_{T}^{0}(\lambda_{1}^{0},\lambda_{2}^{0})$ and $\overline{{\cal S}}_{T}(\tau_{0};\lambda_{2}^{0})$ are given by $1-\Phi(q_{\alpha}-\psi^{0})$ 
and $1-\Phi(q_{\alpha}-\overline{\psi})$ respectively, where 
\begin{ceqn}
\begin{align}
\psi^{0} & =   \left[
\frac{\sqrt{1-\pi_{0}}}{\sigma \sqrt{v^{0}(\lambda_{1}^{0},\lambda_{2}^{0})}} {\bm \gamma}'({\bm Q}_{22}-{\bm Q}_{21}{\bm Q}_{11}^{-1}{\bm Q}_{12}){\bm \gamma} \right], 
\label{eq:17} \\
\overline{\psi} & =  \left[ \frac{\sqrt{1-\pi_{0}}}{\sigma \sqrt{\bar{v}(\tau_{0};\lambda_{2}^{0})}} 
{\bm \gamma}'({\bm Q}_{22}-{\bm Q}_{21}{\bm Q}_{11}^{-1}{\bm Q}_{12}){\bm \gamma} \right],
\label{eq:18}
\end{align}
\end{ceqn}
with $v^{0}(\lambda_{1}^{0},\lambda_{2}^{0})$ and $\bar{v}(\tau_{0};\lambda_{2}^{0})$ as in (\ref{eq:11}) and (\ref{eq:13})-(\ref{eq:14}) and the ${\bm Q}_{ij}$'s referring to the components of 
the population moment matrix ${\bm Q}$ in assumption B1(i). 

}

\vspace{0.2cm}

We note that power is monotonic 
in the sense that both $\psi^{0}$ and $\overline{\psi}$ are non-decreasing as $||{\bm \gamma}||$ gets large. For a given significance level, the larger the two non-centrality parameters are the greater the associated probabilities of rejecting the null hypothesis. 

The expressions in (\ref{eq:17})-(\ref{eq:18}) are particularly useful for highlighting the factors that influence power by shifting the center of the null asymptotic standard normal distributions away from zero.  
Viewing the asymptotic local power functions $\Phi(\psi^{0}-q_{\alpha})$ and $\Phi(\overline{\psi}-q_{\alpha})$ in Corollary 1 as providing approximations to the correct decision frequencies of the two test statistics under a sufficiently large $T$ and {\emph specific alternatives}, we note that for a given size $\alpha$ both test statistics are expected to exhibit a stronger ability to detect departures from the null when the variances of the omitted predictors are large and their correlation with the included predictors small.
 This feature is particularly important since it hints at the fact that the presence of nearly integrated predictors may help enhance power, a scenario we formally consider further below. 

To highlight these points with greater clarity it is useful to focus on the simplified case of two centered predictors ${\bm x}_{t}=(x_{1,t},x_{2.t})$ so that (\ref{eq:17})-(\ref{eq:18}) simplify as
\begin{ceqn}
\begin{align}
\psi^{0} & =  \left[ \frac{\sqrt{1-\pi_{0}}}{\sqrt{v^{0}(\lambda_{1}^{0},\lambda_{2}^{0})}}
\frac{\gamma^{2}}{\sigma}(1-\rho_{12}^{2})E[x_{2,t}^{2}] \right]\label{eq:19}
\end{align}
\end{ceqn}
\begin{ceqn}
\begin{eqnarray}
\overline{\psi} & = \left[ \frac{\sqrt{1-\pi_{0}}}{\sqrt{\bar{v}(\tau_{0};\lambda_{2}^{0})}}
\frac{\gamma^{2}}{\sigma}(1-\rho_{12}^{2})E[x_{2,t}^{2}] \right] \label{eq:20}
\end{eqnarray}
\end{ceqn}
with $\rho_{12}=Corr[x_{1,t},x_{2,t}]$. All other things being equal, power is expected to deteriorate under a noisy omitted predictor that has low variance (low $E[x_{2,t}^{2}]$) and/or that is highly correlated with the included predictor (e.g. $|\rho_{12}| \approx 1$). Interestingly, this also suggests that an ideal setting in terms of power implications is one where omitted predictors are highly persistent 
while included predictors are stationary so that $\rho_{12}\approx 0$ with $E[x_{2,t}^{2}]$ large. 
Another important factor affecting power is the variance of the $u_{t}^{2}$'s which impacts the magnitudes of $\psi^{0}$ and $\overline{\psi}$ via $\sigma \equiv \sqrt{V[u_{t+1}^{2}]}$. Within an NID errors setting for instance we have 
$V[u_{t+1}^{2}]=E[u_{t+1}^{4}]-\sigma^{4}_{u}=2 \sigma^{4}_{u}$ so that {\it all other things being equal}, an environment with high kurtosis will have a detrimental impact on the power properties of both test statistics.\\

\noindent
{\it \textbf{Power enhancing choices for $(\lambda_{1}^{0},\lambda_{2}^{0})$ and $(\tau_{0};\lambda_{2}^{0})$}} \\ 

The non-centrality parameters in (\ref{eq:17})-(\ref{eq:18}) are also useful for 
assessing the impact of $(\lambda_{1}^{0}, \lambda_{2}^{0})$ and $(\tau_{0},{\lambda}_{2}^{0})$ on both the absolute and relative local powers of the two tests and for providing useful guidance on suitable 
choices for those user-inputs.
From Corollary 1, since the mapping $m \mapsto P[Z>q_{\alpha}-m]$ is increasing in $m$ on $[0,\infty)$, a test of size $\alpha$ based on $S_{T}^{0}(\lambda_{1}^{0},\lambda_{2}^{0})$ will be preferable, in terms of its local power, to a test of the same size based on $S_{T}^{0}({\lambda_{1}^{0}}',{\lambda_{2}^{0}}')$
whenever $\psi^{0}(\lambda_{1}^{0},\lambda_{2}^{0})>\psi^{0}({\lambda_{1}^{0}}',{\lambda_{2}^{0}}')$, holding all other parameters entering $\psi^{0}$ constant.
Given $\psi^{0}$ in (\ref{eq:17}) with $v^{0}(\lambda_{1}^{0},\lambda_{2}^{0})$ defined as in (\ref{eq:11})
it follows that those two parameters should be set near their boundary of 1 and in close vicinity of one another (e.g. ${\cal S}_{T}^{0}(\lambda_{1}^{0}=1,\lambda_{2}^{0})$ for $\lambda_{2}^{0}\approx 0.9$ as a possibility). \\

Regarding the average based statistic  $\overline{{\cal S}}_{T}(\tau_{0};\lambda_{2}^{0})$ we note from (\ref{eq:13})-(\ref{eq:14}) that $\overline{\psi}$ in (\ref{eq:18}) viewed as a function of $\lambda_{2}^{0}$ and $\tau_{0}$ (holding all other parameters constant) reaches its unique maximum for 
\begin{ceqn}
\begin{align}
\lambda_{2}^{0} & =  0.5 \ \tau_{0}+0.5  
\label{eq:21}
\end{align}
\end{ceqn}
\noindent
supporting the use of  $\overline{{\cal S}}_{T}(\tau_{0};\lambda_{2}^{0}=0.5\tau_{0}+0.5)$ in its practical implementation. If $\tau_{0}=0.5$ for instance, which corresponds to a test statistic that averages across the largest half of the $\ell_{1}$ magnitudes, this approximate asymptotic power based metric
points to an implementation based on $\overline{{\cal S}}_{T}(\tau_{0}=0.5;\lambda_{2}^{0}=0.75)$. 
Since $\overline{\psi}$ is also a monotonically increasing function of $\tau_{0}$ however, it also 
follows that the same
average based statistic should be operationalised 
with a choice of $\tau_{0}$ that is in the vicinity of 1 (e.g. 
 $\overline{{\cal S}}_{T}(\tau_{0}=0.8;\lambda_{2}^{0}=0.5(0.8)+0.5)$ or 
 $\overline{{\cal S}}_{T}(\tau_{0}=0.9;\lambda_{2}^{0}=0.5(0.9)+0.5)$ as possibilities). A practical side to this power enhancing choice of $\lambda_{2}^{0}$ is that the implementation of $\overline{\cal S}_{T}(\tau_{0};\lambda_{2}^{0})$ essentially requires only a single user-input. \\

Given these preferred parameterisations of the two test statistics it is also useful to evaluate whether either of the two statistics is expected to dominate the other in the sense of $\psi^{0}$ being greater or smaller than $\overline{\psi}$ over particular regions of the pairs $(\lambda_{1}^{0},\lambda_{2}^{0})$ and $(\tau_{0},\lambda_{2}^{0}=0.5\tau_{0}+0.5)$, holding all other parameters constant. Given the standard normal asymptotics of both test statistics a useful metric for comparing their local powers is Pitman's Asymptotic Relative Efficiency (ARE) which here takes particularly simple forms, following directly from Corollary 1. To avoid confusion between the $\lambda_{2}^{0}$ parameter used in ${\cal S}_{T}^{0}(\lambda_{1}^{0},\lambda_{2}^{0})$ and $\lambda_{2}^{0}$ used in $\overline{\cal S}_{T}(\tau_{0},\lambda_{2}^{0})$ we write the two statistics as ${\cal S}_{T}^{0}(\lambda_{1}^{0},\lambda_{2}^{0})$ 
and $\overline{\cal S}_{T}(\tau_{0};\overline{\lambda}_{2}^{0})$ with $\overline{\lambda}_{2}^{0}=0.5 \tau_{0}+0.5$ as in (\ref{eq:21}).
Their ${\text{ARE}}$ is now given by
\begin{ceqn}
\begin{align}
{\text{ARE}}({\cal S}^{0},\overline{\cal S}) & =  \left[\dfrac{\overline{v}(\tau_{0};0.5\tau_{0}+0.5)}{v^{0}(\lambda_{1}^{0};\lambda_{2}^{0})}\right]
\label{eq:22}
\end{align}
\end{ceqn} 
and more specifically
\begin{ceqn} 
\begin{align}
{\text{ARE}}({\cal S}^{0},\overline{\cal S}) & =  \dfrac{\lambda_{1}^{0}\lambda_{2}^{0}}{|\lambda_{1}^{0}-\lambda_{2}^{0}|}
\dfrac{2(1-\tau_{0})(1+\ln((1+\tau_{0})/2))+2\tau_{0}\ln \tau_{0}}{(1-\tau_{0})^{2}}.
\label{eq:23}
\end{align}
\end{ceqn}
From (\ref{eq:23}) we can observe a clear trade-off between $\tau_{0}$ and the magnitudes of $\lambda_{1}^{0}$ and $\lambda_{2}^{0}$
used in  ${\cal S}_{T}^{0}(\lambda_{1}^{0},\lambda_{2}^{0})$. If we focus on $\lambda_{1}^{0}=1$ it follows from (\ref{eq:23}) that 
${\text{ARE}}\geq 1$ for 
\begin{ceqn}
\begin{align}
\lambda_{2}^{0} & \geq \dfrac{(1-\tau_{0})^{2}}{(1-\tau_{0})(3-\tau_{0})+2[\ln 0.5(1+\tau_{0})-\tau_{0}\ln ((1+\tau_{0})/2\tau_{0})]} \label{eq:24}
\end{align}
\end{ceqn}
 which is a monotonically increasing function of $\tau_{0}$ and highlights the fact that the average based statistic
 will dominate ${\cal S}_{T}^{0}(\lambda_{1}^{0}=1,\lambda_{2}^{0})$ in terms of its local power (i.e. ${\text{ ARE}}
 <1$) unless impractically large magnitudes of $\lambda_{2}^{0}$ are used in its implementation. If the average based statistic is implemented with $\tau_{0}=0.8$ for instance, its power properties will dominate ${\cal S}_{T}^{0}(\lambda_{1}^{0}=1,\lambda_{2}^{0})$ unless  $\lambda_{2}^{0}>0.9798$. If it is implemented with $\tau_{0}=0.9$ the average based statistic will again dominate ${\cal S}_{T}^{0}(\lambda_{1}^{0}=1,\lambda_{2}^{0})$ unless $\lambda_{2}^{0}>0.9908$. These values suggest that the average based statistic with 
$\tau_{0}$ set in the vicinity of unity (e.g. $\overline{\cal S}(\tau_{0}=0.8;\overline{\lambda}_{2}^{0}=0.9)$) will dominate ${\cal S}^{0}_{T}(\lambda_{1}^{0}=1,\lambda_{2}^{0})$ in terms of its 
local power unless impractically large magnitudes of $\lambda_{2}^{0}$ are used in ${\cal S}^{0}_{T}(\lambda_{1}^{0}=1,\lambda_{2}^{0})$.

\subsection{Consistency and Local Power under Persistence}

We now consider an environment where the $p$ predictors ${\bm x}_{t}$ entering (\ref{eq:1})-(\ref{eq:2}) are modelled as local to unit root processes specified as 
\begin{ceqn}
\begin{align}
{\bm x}_{t} & = \left({\bm I}_{p}-\dfrac{\bm C}{T}\right){\bm x}_{t-1}+\epsilon_{t} 
\label{eq:25}
\end{align}
\end{ceqn}
\noindent where ${\bm C}=diag(c_{1},\ldots,c_{p})$ for $c_{i}>0$, $i=1,\ldots,p$ and $\epsilon_{t}$ some stationary and ergodic random disturbance process. The new set of assumptions under which we establish our results are now summarised in Assumption B2 below where ${\bm J}_{C}(s)=({\bm J}_{1C}(s),{\bm J}_{2C}(s))'$ denotes a p-dimensional Ornstein-Uhlenbeck process whose two components ${\bm J}_{1C}(s)$ and ${\bm J}_{2C}(s)$ are associated with the dynamics of ${\bm x}_{1,t}$ and ${\bm x}_{2,t}$ respectively. \\

\noindent
ASSUMPTION B2: \\

{\em 
\noindent
(i) $\left(\dfrac{{\bm x}_{[Ts]}}{\sqrt{T}},\dfrac{\sum_{t=1}^{[Ts]} u_{t}}{\sqrt{T}},
		\dfrac{\sum_{t=1}^{[Ts]} (u_{t}^{2}-\sigma^{2}_{u})}{\sqrt{T}}\right) \stackrel{\cal{D}}\rightarrow ({\bm J}_{C}(s),\sigma_{u} W_{u}(s), \sigma W(s))$, $s \in [0,1]$.  \vspace{0.34cm}
	
\noindent	
(ii) Assumption A(iii) holds. \vspace{0.34cm}

\noindent
(iii) The user-inputs in ${\cal S}_{T}(\lambda_{1}^{0},\lambda_{2}^{0})$ and 
$\overline{\cal S}_{T}(\tau_{0};\lambda_{2}^{0})$ are such that 
$(\lambda_{1}^{0},\lambda_{2}^{0})\in \Lambda^{0}$ and $(\tau_{0},\lambda_{2}^{0})\in \overline{\Lambda}^{0}$ respectively. \vspace{0.34cm}
	
}

\noindent
The asymptotic power properties of ${\cal S}_{T}^{0}(\lambda_{1}^{0},\lambda_{2}^{0})$ and $\overline{{\cal S}}_{T}(\tau_{0};\lambda_{2})$ are now summarised in Proposition 4 below. \\

\noindent 
{\bf PROPOSITION 4}.  {\it (i) Suppose model (\ref{eq:2}) holds with ${\bm \beta}_{2}\neq 0$ and fixed, then under Assumption B2 and as $T \rightarrow \infty$ we have 
	$ {\cal S}_{T}^{0}(\lambda_{1}^{0},\lambda_{2}^{0}) \stackrel{p}\rightarrow \infty$ and $\overline{{\cal S}}_{T}(\tau_{0};\lambda_{2}) \stackrel{p}\rightarrow \infty$. (ii)  Suppose model (\ref{eq:2}) holds with ${\bm \beta}_{2}={\bm \gamma}/T^{3/4}$ for ${\bm \gamma\neq 0}$. Under Assumption B2, $\lim_{||{\bm \gamma}||\rightarrow \infty}\lim_{T \rightarrow \infty}{\cal S}^{0}_{T}(\lambda_{1}^{0},\lambda_{2}^{0})=\infty$ and $\lim_{||{\bm \gamma}||\rightarrow \infty}\lim_{T \rightarrow \infty} \overline{{\cal S}}_{T}(\tau_{0};\lambda_{2})=\infty$ in probability.} \\

\noindent
A key message that is conveyed by Proposition 4 when contrasted with Proposition 3 is the important impact of persistence on the power properties the test statistics. The presence of persistent predictors leads to a faster divergence rate for both statistics as reflected in the faster convergence rate towards zero of ${\bm \beta}_{2}$ that can be 
accommodated. With highly persistent predictors, both test statistics diverge at the same $T^{3/2}$ rate compared with a rate of $T^{1/2}$ when predictors were stationary.  

A more explicit formulation of 
the departure from the null distribution in this local to unit-root context can also be highlighted
through the following formulations of the limiting distributions of the two test statistics under the local alternative of interest. \\

\noindent
{\bf COROLLARY 2}. {\it  Suppose model (\ref{eq:2}) holds with ${\bm \beta}_{2}={\bm \gamma}/T^{3/4}$ for ${\bm \gamma\neq 0}$. Under assumption B2 and as $T \rightarrow \infty$ we have
\begin{ceqn}
\begin{align}
{\cal S}_{T}^{0}(\lambda_{1}^{0},\lambda_{2}^{0}) & \stackrel{\cal{D}}\rightarrow  \mathcal{N} (0,1)+{\xi}^{0} \label{eq:26} \\
\overline{{\cal S}}_{T}(\tau_{0};\lambda_{2}) & \stackrel{\cal{D}}\rightarrow 
\mathcal{N}(0,1)+\overline{\xi} \label{eq:27}
\end{align}
\end{ceqn}
with 
\begin{ceqn}
	{\small
\begin{align}
{\xi}^{0} & =  
\frac{\sqrt{1-\pi_{0}}}{\sigma\sqrt{v^{0}(\lambda_{1}^{0},\lambda_{2}^{0})}} 
\ {\bm \gamma}'\left( \frac{1}{(1-\pi_{0})\lambda_{1}^{0}} \int_{\pi_{0}}^{\pi_{0}+(1-\pi_{0})\lambda_{1}^{0}} {\bm J_{C}}^{*}(s) {\bm J_{C}}^{*}(s)'\right) {\bm \gamma}, \label{eq:28} \\
\overline{\xi} & = 
\frac{\sqrt{1-\pi_{0}}}{\sigma \sqrt{\overline{v}(\tau_{0};\lambda_{2}^{0})}}
\ {\bm \gamma}' \left(\frac{1}{(1-\tau_{0})} \int_{\tau_{0}}^{1}  \frac{1}{(1-\pi_{0}) \lambda_{1}}
\left(\int_{\pi_{0}}^{\pi_{0}+(1-\pi_{0})\lambda_{1}} {\bm J_{C}}^{*}(s) {\bm J_{C}}^{*}(s)'\right) d\lambda_{1}   \right){\bm \gamma} \label{eq:29}
\end{align}
}
\end{ceqn}
where ${\bm J}_{C}^{*}(s)={\bm J}_{2C}(s)-{\bm M}(s) {\bm J}_{1C}(s)$ and 
${\bm M}(s)=(\int_{0}^{s}{\bm J}_{1C}{\bm J}_{1C}')^{-1}(\int_{0}^{s}{\bm J}_{1C}{\bm J}_{2C}')$.
}
\\

\noindent 
It is here interesting to compare (\ref{eq:28})-(\ref{eq:29}) with the non-centrality parameters  (\ref{eq:17})-(\ref{eq:18}) obtained in 
the stationary context. The two pairs are essentially analogous in the sense that the constant population moments of predictors (i.e. the ${\bm Q}_{i,j}$'s) are now replaced by stochastic integrals in Ornstein-Uhlenbeck processes (i.e. ${\bm J}_{i,C}$). 
The homogeneity throughout the sample of the limit moment matrix in Assumption B1(i) is of course no longer valid in the 
context of the stochastic integrals in (\ref{eq:28})-(\ref{eq:29}). 
The above results also imply that the role played by the pairs $(\lambda_{1}^{0}, \lambda_{2}^{0})$ and $(\tau_{0},\lambda_{2}^{0})$ in this local to unit-root context will mirror our earlier analysis based on a stationary setting, supporting the same practical implementation of both test statistics in terms of their parameterisations i.e. the power enhancing choices for  $(\lambda_{1}^{0}, \lambda_{2}^{0})$ and $(\tau_{0},\lambda_{2}^{0})$ discussed above continue to hold in the current context.

\section{Power Enhancements}

Here we explore a particular adjustment that can be applied to our two test statistics ${\cal S}_{T}(\lambda_{1}^{0},\lambda_{2}^{0})$ and $\overline{\cal S}_{T}(\tau_{0};\lambda_{2}^{0})$ with the purpose of boosting their asymptotic local power properties without affecting their limiting null distributions. The theoretical principle underlying our proposed approach mirrors the idea in \cite{fly2015} where the authors proposed to augment Wald type statistics with a component that vanishes asymptotically under the null while diverging under alternatives of interest. More formally we seek to augment our proposed two test statistics as 
\begin{align}
{\cal S}_{T,adj}^{0}(\lambda_{1}^{0},\lambda_{2}^{0}) & \equiv {\cal S}_{T}^{0}(\lambda_{1}^{0},\lambda_{2}^{0})+h_{T}^{0}(\lambda_{1}^{0},\lambda_{2}^{0}) \label{eq:30} \\
\overline{{\cal S}}_{T,adj}(\tau_{0};\lambda_{2}^{0}) & \equiv   
\overline{{\cal S}}_{T}(\tau_{0};\lambda_{2}^{0})+\overline{h}_{T}(\tau_{0};\lambda_{2}^{0})
\label{eq:31}
\end{align}
for some suitably chosen $h_{T}^{0}(\lambda_{1}^{0},\lambda_{2}^{0})$ and $\overline{h}_{T}(\tau_{0};\lambda_{2}^{0})$ terms which are such that these adjusted versions of our two test statistics maintain the same limiting null distributions as in Proposition 1 while at the same time displaying more favorable power properties. 

In what follows we show that a particular transformation of the forecast errors $\hat{e}_{2,t+1}$ associated with the larger forecasting model can be used to design such augmentation terms in a way that fulfils the requirement that $h_{T}^{0}(\lambda_{1}^{0},\lambda_{2}^{0}) $ and $\overline{h}_{T}(\tau_{0};\lambda_{2}^{0})$ vanish asymptotically under the null while diverging at a desirable rate under the alternative. The augmentation we propose to consider is motivated by the 
well known Clark and West adjustment to DM type statistics introduced in \cite{cw2007}. The original motivation behind Clark and West's adjustment relied on the intuition that 
under the null hypothesis estimation noise contaminates the ${\hat{e}_{2,t+1}}$'s due to the estimation of 
parameters that are zero in the population. This in turn translates into an inflated $\text{MSE}_{2}$ resulting in 
test statistics that are severely undersized. Clark and West proposed to correct for such distortions by suitably adjusting the magnitudes of the forecast errors estimated from the larger model. 

To lay down the context and with no loss of generality it is useful to operate within a simplified version of (\ref{eq:1})-(\ref{eq:2}), setting ${\bm \delta}_{1}=0$ and ${\bm \beta}_{1}=0$ so that 
$\hat{e}_{1,t+1}=u_{t+1}$ and $\hat{e}_{2,t+1}=u_{t+1}-{\bm x}_{2,t}'(\hat{\bm \beta}_{2}-\bm \beta_{2})$. We can now write $Z_{T}(\ell_{1},\ell_{2})$ in (\ref{eq:6}) as
\begin{ceqn}
	{\small 
	\begin{align}
	\frac{\sum\limits_{t=k_{0}}^{k_{0}+\ell_{1}-1}\hat{e}_{1,t+1}^{2}}{\sqrt{T-k_{0}}}-
	\frac{\ell_{1}}{\ell_{2}} \frac{\sum\limits_{t=k_{0}}^{k_{0}+\ell_{2}-1}\hat{e}_{2,t+1}^{2}}{\sqrt{T-k_{0}}}
	& =  \frac{\sum\limits_{t=k_{0}}^{k_{0}+\ell_{1}-1}u_{t+1}^{2}}{\sqrt{T-k_{0}}}-\frac{\ell_{1}}{\ell_{2}}
	\frac{\sum\limits_{t=k_{0}}^{k_{0}+\ell_{2}-1}u_{t+1}^{2}}{\sqrt{T-k_{0}}} \nonumber \\
	& 
	+2\frac{\ell_{1}}{\ell_{2}} \frac{\sum\limits_{t=k_{0}}^{k_{0}+\ell_{2}-1}(\hat{\bm \beta}_{2,t}-\bm \beta_{2})' {\bm x}_{2,t}u_{t+1}}{\sqrt{T-k_{0}}}
	\nonumber \\
	& -  
	\frac{\ell_{1}}{\ell_{2}} \frac{\sum\limits_{t=k_{0}}^{k_{0}+\ell_{2}-1}(\hat{\bm \beta}_{2,t}-\bm \beta_{2})' {\bm x}_{2,t} {\bm x}_{2,t}'(\hat{\bm \beta}_{2,t}-\bm \beta_{2})}{\sqrt{T-k_{0}}}.
	\label{eq:32}
	\end{align} 
}
\end{ceqn}

Although it is implicit in our Assumption A(i) that the last two terms in the right hand side of (\ref{eq:32}) vanish asymptotically under the null hypothesis, in finite samples the rightmost quadratic form is likely to pull down the spread in MSEs causing their 
null distribution to be mis-centered. Noting that $(\hat{\bm \beta}_{2,t}-\bm \beta_{2})' {\bm x}_{2,t} {\bm x}_{2,t}'(\hat{\bm \beta}_{2,t}-\bm \beta_{2}) \equiv (\hat{e}_{1,t+1}-\hat{e}_{2,t+1})^{2}$,
\cite{cw2007}'s proposal was to reformulate the sample MSE spreads between Models 1 and 2 with 
an adjusted version of $\hat{e}_{2,t+1}^{2}$, say $\widetilde{e}_{2,t+1}^{2}$, given by
\begin{ceqn}
\begin{align}
\widetilde{e}_{2,t+1}^{2} & = \hat{e}_{2,t+1}^{2}-(\hat{e}_{1,t+1}-\hat{e}_{2,t+1})^{2}.
\label{eq:33}
\end{align} 
\end{ceqn}

It turns out that implementing the adjustment in 
(\ref{eq:33}) within our two test statistics (i.e., using $\widetilde{e}_{2,t+1}^{2}$ instead of $\hat{e}_{2,t+1}^{2}$ in 
${\cal S}_{T}^{0}(\lambda_{1}^{0},\lambda_{2}^{0})$ and $\overline{\cal S}_{T}(\tau_{0};\lambda_{2}^{0})$) allows us to reformulate them as in (\ref{eq:30})-(\ref{eq:31}) with $h_{T}^{0}(\lambda_{1}^{0},\lambda_{2}^{0})$ 
and $\overline{h}_{T}(\tau_{0};\lambda_{0}^{2})$ fulfilling the desirable requirements in \cite{fly2015} in the sense that the adjustments do not alter the asymptotic null distributions of the test statistics  while at the same time leading to an increase in the associated non-centrality parameters under the local alternatives of interest. 

The expression in (\ref{eq:32}) is also useful for highlighting what distinguishes 
our framework that operates under $\ell_{1}\neq \ell_{2}$ with a standard approach that sets $\ell_{1}=\ell_{2}=T-k_{0}$ as for instance in all Diebold-Mariano type statistics. 
Under $\ell_{1}=\ell_{2}$ we note that the first two terms in the right hand side of (\ref{eq:32}) cancel out so that the asymptotic behavior of the expression is determined by the two rightmost quadratic forms whose {\it non-normalised} versions have been shown to be $O_{p}(1)$ with non-standard limits (see \cite{cm2001, cm2005}). Allowing $\ell_{1}\neq \ell_{2}$ essentially forces the asymptotics of the MSE spreads to be driven solely by the first two components in the right hand side of (\ref{eq:32}). 

Letting ${\cal S}_{T,adj}^{0}(\lambda_{1}^{0},\lambda_{2}^{0})$
and $\overline{{\cal S}}_{T,adj}(\tau_{0};\lambda_{2}^{0})$ denote the adjusted versions of our two test statistics
it immediately follows from (\ref{eq:33}) and standard algebra that 
\begin{ceqn}
\begin{align}
{\cal S}_{T,adj}^{0}(\lambda_{1}^{0},\lambda_{2}^{0}) & = {\cal S}_{T}^{0}(\lambda_{1}^{0},\lambda_{2}^{0}) 
+\frac{1}{\hat{\sigma}}\frac{1}{\lambda_{2}^{0} \sqrt{v^{0}(\lambda_{1}^{0},\lambda_{2}^{0})}} 
\dfrac{\sum_{t=k_{0}}^{k_{0}+[(T-k_{0})\lambda_{2}^{0}]} (\hat{e}_{1,t+1}-\hat{e}_{2,t+1})^{2}}{\sqrt{T-k_{0}}} \nonumber
\\
& \equiv  {\cal S}_{T}^{0}(\lambda_{1}^{0},\lambda_{2}^{0}) + h_{T}^{0}(\lambda_{1}^{0},\lambda_{2}^{0}) 
\label{eq:34} 
\end{align}
\end{ceqn}
and 
\begin{ceqn}
\begin{align}
\overline{\cal S}_{T,adj}(\tau_{0};\lambda_{2}^{0}) & = \overline{\cal S}_{T}(\tau_{0};\lambda_{2}^{0}) 
+\frac{1}{\hat{\sigma}}\frac{1}{\lambda_{2}^{0} \sqrt{\overline{v}(\tau_{0};\lambda_{2}^{0})}} 
\dfrac{\sum_{t=k_{0}}^{k_{0}+[(T-k_{0})\lambda_{2}^{0}]} (\hat{e}_{1,t+1}-\hat{e}_{2,t+1})^{2}}{\sqrt{T-k_{0}}} \nonumber \\
 & \equiv \overline{\cal S}_{T}(\tau_{0};\lambda_{2}^{0})+\overline{h}_{T}(\tau_{0};\lambda_{2}^{0}).
\label{eq:35}
\end{align}
\end{ceqn}

The expressions in (\ref{eq:34}) and (\ref{eq:35}) highlight the fact that the adjustment to the MSE of the larger model results in test statistics that are augmented versions of ${\cal S}_{T}(\lambda_{1}^{0},\lambda_{2}^{0})$ and $\overline{\cal S}_{T}(\tau_{0};\lambda_{2}^{0})$. As we establish formally below the presence of the additional terms $h_{T}^{0}(\lambda_{1}^{0},\lambda_{2}^{0})$
and $\overline{h}_{T}(\tau_{0};\lambda_{2}^{0})$ 
leaves the limiting null distributions unchanged as both quantities vanish asymptotically. Under the alternative both $h_{T}^{0}(\lambda_{1}^{0},\lambda_{2}^{0})$
and $\overline{h}_{T}(\tau_{0};\lambda_{2}^{0})$ diverge to infinity at the same rate as ${\cal S}_{T}(\lambda_{1}^{0},\lambda_{2}^{0})$ and $\overline{\cal S}_{T}(\tau_{0};\lambda_{2}^{0})$ implying that 
${\cal S}_{T,adj}(\lambda_{1}^{0},\lambda_{2}^{0})$ and $\overline{\cal S}_{T,adj}(\tau_{0};\lambda_{2}^{0})$ will also share the consistency and local power characteristics of their unadjusted counterparts in the sense of 
diverging to infinity as $||{\bm \gamma}||\rightarrow \infty$. 
More importantly however, the presence of $h_{T}^{0}(\lambda_{1}^{0},\lambda_{2}^{0})$
and $\overline{h}_{T}(\tau_{0};\lambda_{2}^{0})$ does result in different (strictly larger) non-centrality parameters that make these adjusted statistics have more favorable power properties. These features are formalised in Proposition 5 and Corollary 3 below. \\

\noindent
{\bf PROPOSITION 5}. {\em The results in Propositions 1-4 continue to hold when ${\cal S}^{0}_{T}(\lambda_{1}^{0},\lambda_{2}^{0})$ and $\overline{{\cal S}}_{T}(\tau_{0};\lambda_{2}^{0})$ are replaced with ${\cal S}_{T,adj}^{0}(\lambda_{1}^{0},\lambda_{2}^{0})$ and $\overline{{\cal S}}_{T,adj}(\tau_{0};\lambda_{2}^{0})$ respectively.
} \\

\noindent
{\bf COROLLARY 3}. {\em (i) Under the assumptions of Corollary 1 (stationary predictors), the asymptotic local power functions of the tests based on ${\cal S}_{T,adj}^{0}(\lambda_{1}^{0},\lambda_{2}^{0})$ and $\overline{{\cal S}}_{T,adj}(\tau_{0};\lambda_{2}^{0})$ 
are given by $1-\Phi(q_{\alpha}-2 \psi^{0})$ and $1-\Phi(q_{\alpha}-2 \overline{\psi})$
with $\psi^{0}$ and $\overline{\psi}$ as in (\ref{eq:17}) and (\ref{eq:18}). 
(ii) Under the assumptions of Corollary 2 (persistent predictors) we have ${\cal \bm S}_{T,adj}^{0}(\lambda_{1}^{0},\lambda_{2}^{0})\stackrel{\cal D}\rightarrow \mathcal{N}(0,1)+\xi_{adj}^{0}$ and $\overline{\cal \bm S}_{T,adj}(\tau_{0};\lambda_{2}^{0})\stackrel{\cal D}\rightarrow \mathcal{N}(0,1)+ \overline{\xi}_{adj}$ where 
\begin{ceqn}
{\small
\begin{align}
\xi_{adj}^{0} & = \xi^{0}+ 
\frac{\sqrt{1-\pi_{0}}}{\sigma \sqrt{v^{0}(\lambda_{1}^{0},\lambda_{2}^{0})}} 
\ {\bm \gamma}' \left(\frac{1}{(1-\pi_{0})\lambda_{2}^{0}}\int_{\pi_{0}}^{\pi_{0}+(1-\pi_{0})\lambda_{2}^{0}} {\bm J_{C}}^{*}(s) {\bm J_{C}}^{*}(s)'\right) {\bm \gamma}, \label{eq:36} \\
\vspace{0.28cm}
\overline{\xi}_{adj} & = \overline{\xi}+ \frac{\sqrt{1-\pi_{0}}}{\sigma \sqrt{\overline{v}(\tau_{0};\lambda_{2}^{0})}} 
\ {\bm \gamma}' \left(\frac{1}{(1-\pi_{0})\lambda_{2}^{0}} \int_{\pi_{0}}^{\pi_{0}+(1-\pi_{0})\lambda_{2}^{0}} {\bm J_{C}}^{*}(s) {\bm J_{C}}^{*}(s)'\right)  {\bm \gamma} \label{eq:37}
\end{align}
}
\end{ceqn}
with $\xi^{0}$ and $\overline{\xi}$ as in (\ref{eq:28}) and (\ref{eq:29}).
} \\

Proposition 5 essentially implies that all our results regarding the null limiting distributions of  ${\cal S}_{T}(\lambda_{1}^{0},\lambda_{2}^{0})$ and $\overline{\cal S}_{T}(\tau_{0};\lambda_{2}^{0})$ and their general power properties (consistency and detectability of local departure from the null) continue to hold for their adjusted counterparts while Corollary 3 documents important differences in their specific non-centrality terms. 

Indeed, the results in Corollary 3 are particularly interesting and useful for the practical assessment of the power properties of the adjusted versus unadjusted statistics. We have an environment whereby the limiting distributions of the two types of test statistics are the same under the null hypothesis while their non-centrality parameters differ under the alternative, pointing to a more favorable behavior for the adjusted statistics when it comes to detecting local departures from the null.

Letting $\psi^{0}_{adj}$ and $\overline{\psi}_{adj}$ denote the non-centrality parameters associated with the adjusted statistics, Corollary 3(i) establishes that in a stationary context we have $\psi^{0}_{adj}=2 \psi^{0}$ and $\overline{\psi}_{adj}=2\overline{\psi}$ so that $\psi^{0}_{adj}/\psi_{0}=2$ and $\overline{\psi}_{adj}/\overline{\psi}=2$. 
In the case of persistent predictors the comparison between $\xi^{0}$ and $\xi^{0}_{adj}$ and between $\overline{\xi}$ and  
$\overline{\xi}_{adj}$ also indicates that the adjusted quantities will stochastically dominate their non-adjusted counterparts 
in the sense that $P[\xi^{0}_{adj}>q]\geq P[\xi^{0}>q]$ and $P[\overline{\xi}_{adj}>q]\geq P[\overline{\xi}>q]$ for some given critical value $q$ and this is again expected to translate into more favorable power outcomes for the adjusted statistics under persistent predictors as well.

\section{Empirical Size and Power}

In this section we investigate the size and power properties of ${\cal S}_{T}^{0}(\lambda_{1}^{0}=1,\lambda_{2}^{0})$ and
$\overline{{\cal S}}_{T}(\tau_{0};\lambda_{2}^{0})$ together with their adjusted versions across two DGPs calibrated to commonly encountered applications and sample sizes in macroeconomics and finance.  
The experiments are designed to emphasise the role  of the pairs ($\lambda_{1}^{0}, \lambda_{2}^{0}$) and $(\tau_{0};\lambda_{2}^{0})$ 
on inferences with the choice of their magnitudes guided by the analysis surrounding our results in Corollaries 1 and 2.  

More specifically, the implementation of ${\cal S}_{T}^{0}(\lambda_{1}^{0},\lambda_{2}^{0})$  is restricted to $\lambda_{1}^{0}=1$ across
$\lambda_{2}^{0} \in \{0.5, 0.55, 0.6, 0.65, 0.7, 0.75, 0.8, 0.85, 0.9, 0.95\}$ and similarly for 
${\cal S}_{T,adj}^{0}(\lambda_{1}^{0}=1,\lambda_{2}^{0})$, thus 
providing a very broad coverage across a range of user-inputs. The average based statistic 
$\overline{{\cal S}}_{T}(\tau_{0};{\lambda}_{2}^{0})$ and its adjusted version $\overline{{\cal S}}_{T,adj}(\tau_{0};{\lambda}_{2}^{0})$ are implemented for $\tau_{0}\in \{0.5, 0.8\}$ across 
${\lambda}_{2}^{0} \in \{0.50, 0.60, 0.70, 0.75, 0.80, 0.85, 0.90, 0.95, 1.00\}$. 

All our size and power simulations below set $\pi_{0}=0.25$ (i.e., $k_{0}=[T \ 0.25]$) to initiate the recursively generated forecasts.

\subsection{DGP1}

A specification that mimics a frequently encountered setting in the 
asset pricing literature is one where the null model is the martingale difference sequence, $y_{t+1}=u_{t+1}$, and the larger model the single predictor based predictive regression, $y_{t+1}=\beta x_{t}+u_{t+1}$ with  $x_{t}=\phi_{1} x_{t-1}+v_{t}$.
Letting $\Sigma=\{\{\sigma^{2}_{u},\rho_{uv}\sigma_{u}\sigma_{v}\},\{\rho_{uv}\sigma_{u}\sigma_{v},\sigma^{2}_{v}\}\}$ denote the covariance of $(u_{t},v_{t})$, in line with commonly encountered magnitudes from the equity premium predictability literature we set $\sigma^{2}_{u}=3$, $\sigma^{2}_{v}=0.01$, $\rho_{uv}=-0.8$ and experiment with $\phi_{1} \in \{0.75,0.95, 0.98\}$. 
The conditionally homoskedastic setting takes $(u_{t},v_{t})\sim NID(0,\Sigma)$ while 
conditional heteroskedasticity is modelled via an ARCH(1) specification, writing 
$u_{t}=\epsilon_{t}\sqrt{h_{t}}$ with $h_{t}=\alpha_{0}+\alpha_{1}u_{t-1}^{2}$ and $\epsilon_{t} \sim NID(0,1)$. This latter choice naturally influences the magnitudes of $\sigma^{2}_{u}$ and $\rho_{uv}$ chosen above and we parameterise $\{\alpha_{0},\alpha_{1}\}$ in a way that maintains the same magnitude for $\sigma^{2}_{u}$ as in the conditionally homoskedastic case i.e. $\alpha_{0}/(1-\alpha_{1})=\sigma^{2}_{u}$. For this purpose we set $(\alpha_{0},\alpha_{1})=(1.8, 0.4)$ throughout. 

Size experiments set $\beta=0$ while for the power properties of the tests we fix the sample size at $T=500$ and evaluate correct decision 
frequencies as $\beta$ moves away from the null with $\beta \in \{0,-1.5,-1.75,-2.0,-2.25, -2.5,-3,-3.5\}$. For
$\beta=\gamma/T^{1/4}$ this is equivalent to $|\gamma|$ increasing  with $\gamma \in \{0,-7.1,-8.3,-9.5,-10.6,-11.8,-14.2,-16.6\}$. 
Lastly, all of the above experiments are conducted using two alternative estimators for $\sigma$. The first one denoted $\hat{\sigma}^{2}_{hom}$ is suitable under conditional homoskedasticity while the second one denoted $\hat{\sigma}^{2}_{nw}$ is its robustified version \`{a} la Newey-West. Both estimators are based on the residuals from the model estimated under the alternative.   

As our Monte-Carlo simulations encompass a very broad range of scenarios and test statistic parameterisations we provide an extensive selection of outcomes in a supplementary online appendix accompanying this paper. Our focus below is on a selection of key size/power results under conditional homoskedasticity and test statistic parameterisations that mainly rely on our recommendations based on our theoretical local power analysis above. \\

\noindent
{\it \textbf{Empirical Size}} \\

Table \ref{tab:Tab1}  presents size estimates for ${\cal S}_{T}^{0}(\lambda_{1}^{0}=1,\lambda_{2}^{0})$ and ${\cal S}_{T,adj}^{0}(\lambda_{1}^{0}=1,\lambda_{2}^{0})$ 
across a broad range of parameterisations in a conditionally homoskedastic setting. For both test statistics 
we note good to excellent matches of the nominal size of 10\% across almost all choices of $\lambda_{2}^{0}$  
for $T\geq 500$. The adjusted statistic ${\cal S}_{T,adj}^{0}(\lambda_{1}^{0}=1,\lambda_{2}^{0})$ in particular has empirical sizes that almost perfectly match 10\% for virtually all magnitudes of $\lambda_{2}^{0}$. 
Under $\phi_{1}=0.75$ for instance, 
${\cal S}_{T,adj}^{0}(\lambda_{1}^{0}=1,{\lambda_{2}^{0}=0.8})$ has resulted in an empirical size of 10.4\% for $T=1000$ and 11.0\% for $T=500$. The corresponding figures for $\phi_{1}=0.95$ were 10.6\% and 11.2\% respectively, thus also highlighting the robustness of the test statistics to the degree of persistence of the predictors as expected from our results in Propositions 1-2. Similar outcomes also characterise ${\cal S}_{T,adj}^{0}(\lambda_{1}^{0}=1,{\lambda_{2}^{0}=0.9})$ suggesting that these test statistics maintain good size control for $T\geq 500$ even when $\lambda_{2}^{0}$ is as large as 0.90 or 0.95 and $\lambda_{1}^{0}$ is set equal to one. 
\begin{center}
	{\bf Table \ref{tab:Tab1}}
\end{center}
\noindent
Regarding the size properties of the unadjusted ${\cal S}_{T}^{0}(\lambda_{1}^{0},\lambda_{2}^{0})$ statistic we note a mild undersizeness for magnitudes of $\lambda_{2}^{0}$ that are in the vicinity of 1, with its empirical sizes clustered around 7-8\%. 
Overall the outcomes in Table \ref{tab:Tab1} have highlighted remarkably stable size properties for both the unadjusted and adjusted statistics across the different magnitudes of $\lambda_{2}^{0}$ including when it is set at 0.90 or 0.95. This is particularly
reassuring given our earlier theoretical power analysis which pointed at desirable parameterisations that satisfy $\lambda_{1}^{0}\approx \lambda_{2}^{0}$ with both $\lambda_{1}^{0}$ and $\lambda_{2}^{0}$ set in the vicinity of 1 in the practical implementation of ${\cal S}_{T}^{0}(\lambda_{1}^{0},\lambda_{2}^{0})$ and ${\cal S}_{T,adj}^{0}(\lambda_{1}^{0},\lambda_{2}^{0})$. 

Before proceeding further it is also useful to briefly rationalise the size behavior of these two statistics when $\lambda_{2}^{0}$ is chosen to lie almost at its boundary as when we set $\lambda_{2}^{0}=0.95$. In such instances we noted the mild undersizeness of 
${\cal S}_{T}^{0}(\lambda_{1}^{0}=1,\lambda_{2}^{0})$ and mild oversizeness of ${\cal S}_{T,adj}^{0}(\lambda_{1}^{0}=1,\lambda_{2}^{0})$ when operating with small to moderately sized samples. A magnitude of $\lambda_{2}^{0}$ that is close to 1 essentially translates into more `MSE content' from the larger model and hence 
a greater exposure to estimation noise when the null model holds true. This results in the unadjusted ${\cal S}_{T}^{0}(\lambda_{1}^{0}=1,\lambda_{2}^{0})$ test statistic's distribution being pushed leftward with fewer than 
expected rejections of the null. On the other hand the adjusted statistic which aims to correct for estimation error via 
(\ref{eq:32}) sees its correction factor's contribution increase as $\lambda_{2}^{0}\rightarrow 1$, a correction factor that is overly inflated in small samples. At this stage it is also useful to point out that the effective sample size is given by $T-k_{0}$ so that under $\pi_{0}=0.25$ and $T=250$ we have only about 188 data points when implementing the tests. A highly persistent predictor combined with such a small sample size can be seen to result in some degree of oversizeness for
$S_{T,adj}(\lambda_{1}^{0},\lambda_{2}^{0})$ based inferences when $|\lambda_{1}^{0}-\lambda_{2}^{0}|$ is particularly small (e.g., for $(\lambda_{1}^{0},\lambda_{2}^{0})=(1,0.95)$). Nevertheless, these finite sample distortions quickly fade away as we increase the sample size to $T=500$.  

For comparison purposes the last column of Table \ref{tab:Tab1} also includes the corresponding size estimates for the DM and CW statistics. These conform with the consensus view that the DM statistic is severely undersized under such nested settings while the CW statistics' empirical sizes are clustered around 5\% for a nominal size of 10\%, in line with the simulation results in Clark and West (2007). 

We next 
consider the finite sample size properties of the average based statistics 
$\overline{{\cal S}}_{T}(\tau_{0};\lambda_{2}^{0})$ and $\overline{{\cal S}}_{T,adj}(\tau_{0};\lambda_{2}^{0})$. 
Recall that the averaging is performed across a portion of the null model's MSE as captured by $\tau_{0}$ and for a given fraction of the second model's MSE $\lambda_{2}^{0}$. Here we present outcomes obtained under $\tau_{0}=0.8$ which only sum across the larger magnitudes of $\lambda_{1}$. Such a choice is theoretically justified by our earlier power analysis with further scenarios presented in the supplementary appendix. Results are presented in Table \ref{tab:Tab2} from which we note that the adjusted statistic $\overline{{\cal S}}_{T,adj}({\tau_{0}=0.8};\lambda_{2}^{0})$ displays good to excellent size control (e.g. empirical size estimates near 10\% under $\lambda_{2}^{0}=1$) under moderate to large sample size choices. 
\begin{center}
	{\bf Table \ref{tab:Tab2}}
\end{center}
\noindent An exception to this is when $\lambda_{2}^{0}\approx 0.5 \ \tau_{0}+0.5 (\approx 0.9 \ here)$ under which it shows a tendency to overreject the null hypothesis in smaller samples.  This is in complete agreement with our earlier theoretical power analysis where we showed that holding all else constant the power of the test statistic must peak under $\lambda_{2}^{0}=0.5 \tau_{0}+0.5$. Thus the empirical sizes peaking for $\lambda_{2}^{0}$ in the vicinity of $0.5(0.8)+0.5=0.90$
highlight the size vs power trade-off that will characterise this average based test statistic. 

Regarding the unadjusted statistic $\overline{{\cal S}}_{T}({\tau_{0}=0.8};\lambda_{2}^{0})$ we can note a tendency to underreject (e.g. empirical sizes in the vicinity of 7\% under $\tau_{0}=0.8$) and this undersizeness deteriorating as $\lambda_{2}^{0} \rightarrow 1$ and $\phi_{1}$ gets closer to 1. This behavior conforms with the intuition that estimation noise caused by the estimation of parameters that are zero in the population 
pushes the test statistic too much to the left, a feature that was the key motivation behind Clark and West's adjustment to the DM statistic. 
Note for instance that these distortions are substantially dampened when the test statistic is implemented with smaller magnitudes of $\lambda_{2}^{0}$ for which it shows good to excellent size control.\\

\noindent
{\it \textbf{Empirical Power}} \\

Table \ref{tab:Tab3} presents empirical power estimates for  ${\cal S}_{T}^{0}(\lambda_{1}^{0}=1,\lambda_{2}^{0})$
and  ${\cal S}_{T,adj}^{0}(\lambda_{1}^{0}=1,\lambda_{2}^{0})$ across $\lambda_{2}^{0}\in \{0.80, 0.85, 0.90, 0.95\}$. The sample size is fixed at $T=500$ and power is evaluated as the DGP moves further away from the null hypothesis.  The choices of $\lambda_{2}^{0}$ are dictated by our theoretical results in Corrollaries 1-2 which pointed to magnitudes satisfying $\lambda_{1}^{0}\approx \lambda_{2}^{0}\approx 1$. For both test statistics we note 
the tendency of their empirical power to converge to 1 as $|\gamma|$ is allowed to increase. We can also clearly observe the particularly favorable impact that the degree of persistence of predictors has on power. As expected from our findings in Propositions 3-4 and their corollaries, power improves as $\lambda_{2}^{0}\rightarrow 1$ and as $\phi_{1}\rightarrow 1$. 
\begin{center}
	{\bf Table \ref{tab:Tab3}}
\end{center}
Given the good overall size control displayed by ${\cal S}_{T,adj}^{0}(\lambda_{1}^{0}=1,\lambda_{2}^{0})$  
Table \ref{tab:Tab3} clearly highlights the benefits of basing inferences on this adjusted version of our first test statistic
calibrated to $\lambda_{1}^{0}=1$ and $\lambda_{2}^{0}\approx 0.9$, also noting that its performance improves considerably as $\phi_{1}\rightarrow 1$. 
For $\beta_{2}=-2$ for instance it displays power in the vicinity of 70\%-80\% under $\phi_{1}=0.75$ and 100\% under $\phi_{1}=0.95$ or $\phi_{1}=0.98$. It is here important to relate our simulation outcomes with the theoretical 
results of Corollary 3 where we established  an ARE of 2 for the adjusted statistic ${\cal S}_{T,adj}^{0}(\lambda_{1}^{0},\lambda_{2}^{0})$ relative to its unadjusted counterpart. This theoretical power enhancement 
is clearly supported by the empirical power estimates of Table \ref{tab:Tab3}. Compare for instance the empirical power of 51.2\% 
for the unadjusted statistic under $\phi_{1}=0.75$ and $\beta=-2.25$ with 81.2\% for its adjusted version, a power gain of 30 percentage points. 

At this stage it is also important to recall that our theoretical analysis based on the asymptotic relative efficiency of  ${\cal S}_{T,adj}^{0}(\lambda_{1}^{0},\lambda_{2}^{0})$ versus 
$\overline{{\cal S}}_{T,adj}({\tau_{0}};\lambda_{2}^{0})$ clearly pointed to the potentially superior power performance of the average based statistic, for larger magnitudes of $\tau_{0}$ in particular. This is clearly corroborated by the comparison between the power outcomes in Table \ref{tab:Tab3} and Table \ref{tab:Tab4} with the latter presenting power outcomes for the $\overline{\cal S}_{T}(\tau_{0}=0.8;\lambda_{2}^{0})$ and $\overline{\cal S}_{T,adj}(\tau_{0}=0.8;\lambda_{2}^{0})$ statistics.  
Focusing on the ``optimal'' choice of $\lambda_{2}^{0}=0.5 \tau_{0}+0.5=0.90$ when $\tau_{0}=0.8$ we note from Table \ref{tab:Tab4} that
$\overline{{\cal S}}_{T,adj}({\bm \tau_{0}=0.8};\lambda_{2}^{0}=0.9)$ clearly dominates all configurations of 
${\cal S}_{T,adj}^{0}(\lambda_{1}^{0},\lambda_{2}^{0})$ in terms of its power properties, typically resulting in relative power gains in excess of 10 percentage points. 
\begin{center}
{\bf Table \ref{tab:Tab4}}
\end{center}

Before proceeding further it is also useful to comment on the power behavior of the DM and CW statistics in comparison to $\overline{{\cal S}}_{T}({\bm \tau_{0}=0.8};\lambda_{2}^{0}=0.90)$. Despite being severely undersized and theoretically unsuitable in the present nested context we note that the DM statistic does show a reasonable ability to detect departures from the null. However we can also observe that it is uniformly dominated by $\overline{{\cal S}}_{T,adj}({\bm \tau_{0}=0.8};\lambda_{2}^{0}=0.90)$ which under $\phi_{1}=0.75$ for instance exceeds its power by about ten percentage points. Comparing the power performance of $\overline{{\cal S}}_{T,adj}({\bm \tau_{0}=0.8};\lambda_{2}^{0}=0.90)$ with that of the CW statistic we note that these two test statistics display very similar power outcomes across most scenarios. Although the CW statistic does not have a well defined limiting distribution due to the nestedness of the competing models it appears to display reasonably good power properties across the DGPs we have considered, despite being far-off the standard normal distribution under the null (as implied by its size properties).

\subsection{DGP2}

The second DGP allows for multiple predictors and is calibrated to mimic US inflation based predictive regressions as considered in \cite{sw2010} and \cite{ghm2014}. We use the same setting as in \cite{ghm2014} and consider a DGP given by $y_{t+1}=\mu+\rho y_{t}+\beta_{1}x_{1,t}+\beta_{2}x_{2,t}+\beta_{3}x_{3,t}+u_{t+1}$ with $\mu=1$ and 
$\rho=0.25$. The predictors ${\bm x}_{t}=(x_{1,t},x_{2,t},x_{3,t})'$ follow the VAR(1) process $\bm x_{t}= \Phi \ \bm x_{t-1}+\bm v_{t}$ 
with ${\Phi}=\{\{0.6,0.1,0\},\{0.6,0.25,0\},\{0,0,0.9\}\}$ thus encompassing both persistent and much noisier processes while also being interdependent. 
The conditionally homoskedastic scenario takes $(u_{t},v_{1,t},v_{2,t},v_{3,t})'\sim NID(0,I_{4})$ while conditional heteroskedasticity is captured via an ARCH(1) process for $u_{t}$ as in DGP1 with $\alpha_{0}=0.6$ and $\alpha_{1}=0.4$ so that its unconditional variance matches unity as in the homoskedastic scenario. 

For our size experiments we set $\beta_{1}=\beta_{2}=\beta_{3}=0$ and our power analysis 
focuses on alternatives to $\beta_{1}=\beta_{2}=\beta_{3}=0$
by fixing  $(\beta_{1},\beta_{2},\beta_{3})=(0.15,0.15,-0.15)$ and evaluating rejection rates of the null hypothesis for $T=250, 500, 1000$. \\

\noindent
{\it \textbf{Empirical Size}} \\

Tables \ref{tab:Tab5} and \ref{tab:Tab6} present empirical size estimates corresponding to the null DGP 
under $\beta_{1}=\beta_{2}=\beta_{3}=0$ for ${\cal S}_{T}^{0}(\lambda_{1}^{0}=1,\lambda_{2}^{0})$ and  
$\overline{{\cal S}}_{T}(\tau_{0}=0.80;\lambda_{2}^{0})$ respectively together with their adjusted versions. As DGP2 contains a larger number of predictors 
than DGP1 we expect the impact of estimation error on the MSE of the second/larger model to be more pronounced 
under the null. This is indeed corroborated by the size estimates in Table \ref{tab:Tab5} where we note the undersizeness of 
the unadjusted ${\cal S}_{T}^{0}(\lambda_{1}^{0},\lambda_{2}^{0})$ statistic which is biased downward and thus results in too 
few rejections of the null (e.g. 6.4\% under $T=1000$ and $\lambda_{2}^{0}=0.9$ versus a nominal size of 10\%). Furthermore, its undersizeness tends to deteriorate for larger magnitudes of $\lambda_{2}^{0}$ as this translates into an increased influence of the larger model's MSE. 
\begin{center}
	{\bf Tables \ref{tab:Tab5}-\ref{tab:Tab6}}
\end{center}
\noindent
The adjusted version of the test statistic  ${\cal S}_{T,adj}^{0}(\lambda_{1}^{0},\lambda_{2}^{0})$
on the other hand is quite effective in adjusting for estimation noise (e.g. the earlier empirical size of 6.4\% is now pushed up to 12\%), while also showing a tendency to ``over-adjust'' in small to moderately size samples, in particular for larger magnitudes of $\lambda_{2}^{0}$. Table \ref{tab:Tab6} presents the corresponding size estimates for the average based statistic $\overline{{\cal S}}_{T,adj}({\bm \tau_{0}=0.8};\lambda_{2}^{0})$. We note that this latter statistic maintains good to excellent size control for moderate sample sizes across all magnitudes of $\lambda_{2}^{0}$ but requires larger samples when $\lambda_{2}^{0}$ is set near one. \\

\noindent
{\it \textbf{Empirical Power}} \\
 
For this DGP our power experiments focus on documenting the rejection frequencies of the null hypothesis for a fixed alternative as the sample size is allowed to increase. Results are presented in Tables \ref{tab:Tab7}-\ref{tab:Tab8} for 
${\cal S}_{T}^{0}(\lambda_{1}^{0},\lambda_{2}^{0})$ and $\overline{{\cal S}}_{T}({\bm \tau_{0}=0.8};\lambda_{2}^{0})$ and their adjusted versions. Under either $T=500$ or $T=1000$ all test statistics have powers at or near 100\%. 
\begin{center}
	{\bf Tables \ref{tab:Tab7}-\ref{tab:Tab8}}
\end{center}
Tables \ref{tab:Tab7}-\ref{tab:Tab8} also clearly corroborate our theoretical power analysis that highlighted peaking powers under 
$\lambda_{2}^{0}=0.5 \tau_{0}+0.5$. Focusing on  $\overline{{\cal S}}_{T}({\bm \tau_{0}=0.8};\lambda_{2}^{0})$ we can clearly observe the empirical powers to be largest under $\lambda_{2}^{0}=0.9$ for all sample sizes (e.g. 98.2\% versus 95.7\% when $\lambda_{2}^{0}=1$ and for $T=250$). \\

The main findings from our simulation experiments can be summarised as follows. 
(i) The adjusted versions of the two test statistics ${\cal S}_{T,adj}^{0}(\lambda_{1}^{0},\lambda_{2}^{0})$ and $\overline{{\cal S}}_{T,adj}({\bm \tau_{0}};\lambda_{2}^{0})$
displayed good to excellent size control across most parameterisations of their respective inputs ($\lambda_{1}^{0}, \lambda_{2}^{0})$ and ($\tau_{0} ,\lambda_{2}^{0})$. (ii) Both test statistics are consistent and have non-trivial local asymptotic power while their finite sample power properties are strongly influenced by the respective magnitudes of those same 
inputs. The guidelines provided by our theoretical power analysis do however lead to highly favorable power
outcomes with implementations such as ${\cal S}_{T,adj}^{0}(\lambda_{1}^{0}=1,\lambda_{2}^{0}\approx 0.9)$ and $\overline{{\cal S}}_{T,adj}({\bm \tau_{0} \approx 0.8};\lambda_{2}^{0}\approx 0.9)$ standing out in terms of their size/power trade-offs, especially for moderately sized samples such as $T \geq 500$. (iii) The proposed methods are valid irrespective of the degree of persistence of the predictors as also corroborated by our finite sample simulations. (iv) Our Monte-Carlo analysis did show that Clark and West's CW statistic which although not grounded on formal standard normal asymptotics also performed particularly 
well in terms of its power, despite relatively important size distortions.
Strictly speaking the CW statistic has been introduced for handling nested models estimated via a rolling as opposed to a recursive approach since from a theoretical standpoint it continues to suffer from the variance degeneracy problem characterising DM type constructions. 

\subsection{Summary and tuning-parameter guidelines}

The above simulation based outcomes combined with our earlier local power analysis 
point to precise guidelines for the choice of tuning parameters required in the implementation of our test statistics. For the ${\cal S}_{T,adj}^{0}(\lambda_{1}^{0},\lambda_{2}^{0})$ statistic we argued that $\lambda_{1}^{0}$ and $\lambda_{2}^{0}$ should be set near their boundary of one and in close vicinity of one another. Our simulations based on ${\cal S}_{T,adj}^{0}(\lambda_{1}^{0}=1,\lambda_{2}^{0})$ for $\lambda_{2}^{0}$ set in the 0.80-0.95 range have indeed resulted in good to excellent size-power tradeoffs and good to excellent size control. 
The robustness of the empirical size outcomes to a much broader range of the $\lambda_{2}^{0}$ magnitudes
is also noteworthy as illustrated by the outcomes in Tables 1 and 5. \\

Regarding the $\overline{\cal S}_{T,adj}(\tau_{0};\lambda_{2}^{0})$ statistic, our theoretical local power analysis led us to argue for $\tau_{0}$ to be set in the vicinity of unity and $\lambda_{2}^{0}$ as $\lambda_{2}^{0}=0.5 \ \tau_{0}+0.5$. 
Simulations based on $\tau_{0}=0.8$ did indeed result in good to excellent size and power properties. As the above choice for $\lambda_{2}^{0}$
is a maximiser of local power it is perhaps natural to expect some size distortions in smaller samples when $\lambda_{2}^{0}$ is set in this way and in particular when 
this is combined with the presence of highly persistent predictors. This is indeed confirmed by our size experiments implemented under $(\tau_{0};\lambda_{2}^{0})=(0.8,0.9)$ and a local-to-unity type predictor. Nevertheless, these specific finite sample distortions can also be seen to progressively vanish as the sample size is allowed to grow.

\section{Application}

We illustrate the implementation of our proposed methods by revisiting a widely considered puzzle in the international economics literature, namely the random walk like behavior of exchange rates. Our goal is to use 
our new test statistics in order to evaluate whether past exchange rate levels have any predictive power for subsequent exchange rate changes. Letting $s_{t}$ denote the log of the spot exchange rate, we compare the out of sample predictive accuracy of the larger model $\Delta s_{t+1} = \alpha + \beta \ s_{t}+u_{t+1}$ (model 2) with the random walk with drift specification $\Delta s_{t+1} = \alpha+u_{t+1}$ (model 1). 

We consider six major currencies (EUR, YEN, GBP, CHF, AUD and CAD) and implement our tests on daily spot rates spanning the period between 1999/01/04 and 2021/07/16, sourced from the Saint-Louis Fred database. An important advantage of the methods developed in this paper is their robustness to the persistence properties of predictors which is particularly relevant when considering exchange rate series. Indeed, for all three daily series we have considered 
the first order autocorrelation coefficient from an AR(1) fit is 0.99.   
\begin{center}
{\bf Table \ref{tab:Tab9}}
\end{center}
Predictive accuracy testing outcomes (p-values) based on ${\cal S}_{T,adj}^{0}(\lambda_{1}^{0},\lambda_{2}^{0})$ and 
$\overline{{\cal S}}_{T,adj}(\tau_{0};\lambda_{2}^{0})$ are presented in Table \ref{tab:Tab9} where we used 
$\pi_{0}=0.5$ to initiate the expanding window estimation (i.e., starting from the middle of the sample). 
For robustness considerations inferences based on ${\cal S}_{T,adj}(\lambda_{1}^{0}=1,\lambda_{2}^{0})$ are implemented across $\lambda_{2}^{0}=\{0.80,0.85,0.90,0.95\}$ while for $\overline{\cal S}_{T,adj}(\tau_{0};\lambda_{2}^{0}=1)$ we consider $\tau_{0}=0.80$ and $\tau_{0}=0.90$. Looking at the top and middle panels of Table \ref{tab:Tab9} we note that results unanimously corroborate the fact that the level of exchange rates does not have any meaningful forecasting power for future currency returns over the period considered and across all major currencies. This result based on the use of daily data also corroborates the recent findings in \cite{ew2021} based on monthly frequencies. The bottom panel of Table \ref{tab:Tab9} displays the p-values associated with the standard DM and the CW statistics. It is here interesting to point out that inferences based on the standard CW statistic lead to a rejection of the random walk specification for the EURUSD and CADUSD series when implementing the test at a 5\% level or above which is in sharp contrast with the large p-values obtained using our two test statistics. 

\vspace{-0.4cm}

\section{Concluding Remarks}

The main motivation of this paper was to provide a way of bypassing the variance degeneracy problem that arises in the context of out-of-sample nested model comparisons. We did so by developing two new test statistics 
shown to have nuisance parameter free standard normal asymptotics and good power properties including in the close vicinity of the null hypothesis. Our proposed inferences can trivially accommodate 
conditional heteroskedasticity and are also shown to be robust to 
the presence of highly persistent predictors. Although our power analysis has ruled out the case of deterministic trends via Assumption B1 for instance, these can also be easily accommodated within our framework without any changes to the implementation of the tests provided that the trends are formulated in a scaled form as $(t/T)$ and its powers. Nested comparisons in purely deterministic environments would be particularly relevant in areas such as temperature modelling (e.g., \cite{wuzhao2017}, \cite{grg2020}) or the recent literature on modelling pandemic dynamics (e.g., \cite{jiangshao2020}, \cite{lilinton2021}). 

Although our proposed test statistics 
require two user inputs each, both have been shown to display good to excellent size control across a very broad range of such parameterisations in a multitude of empirically relevant settings. Although these user-inputs do have considerable influence on the finite sample power properties of both test statistics their choices can be accurately guided by examining the power functions associated with each test statistic, as demonstrated by our simulations and their theoretical backing.   

It is important to recognise that our proposed test statistics do involve the discarding of some information 
albeit very limited and with its amount under the control of the user. As a result some power loss is of course unavoidable but the absence of any alternative approach that uses more information while achieving the same purpose 
in an environment that can accommodate both stationary and persistent predictors as well as conditional heteroskedasticity makes such power losses only notional. 
Our simulation results have indeed shown that very little needs to be discarded for our methods to work well and to provide reliable inferences. In this sense they are not subject to the 
disadvantages of sample splitting based techniques used for instance in the goodness of fit literature (e.g. half-sample methods). 

The principles underlying our proposed inferences based on (\ref{eq:6}) should also be portable beyond out of sample forecasting considerations to areas involving model selection testing \`{a} la \cite{v1989} where nestedness versus non-nestedness or the overlapping nature of models being compared influences test procedures due to variance degeneracy problems (see also \cite{s2015}). In \cite{sw2017} for instance, the authors developed a model selection test
for choosing between two parametric likelihoods based on sample splitting principles which although different from our approach 
based on MSE comparisons on overlapping intervals was driven by similar concerns. Adapting the analysis of this paper to such model selection testing contexts is a promising avenue currently being explored.

\newpage
\begin{center}
{\bf APPENDIX A: TABLES}
\end{center}

\setcounter{table}{0}
\renewcommand{\thetable}{\arabic{table}}

\begin{table}[htbp]
		\centering
	\tabcolsep=0.2cm
	\caption{\footnotesize {\bf DGP1} Empirical size of ${\cal S}_{T}^{0}(\lambda_{1}^{0}=1,\lambda_{2}^{0})$ and ${\cal S}_{T,adj}^{0}(\lambda_{1}^{0}=1,\lambda_{2}^{0})$ under conditional homoskedasticity and 10\% nominal size}
	\vspace{0.2cm}
\scalebox{0.78}{	\begin{tabular}{llccccccccccc}
		& $\lambda_{2}^{0}$ & 0.500 & 0.550 & 0.600 & 0.650 & 0.700 & 0.750 & 0.800 & 0.850 & 0.900 & 0.950 &    \\ \hline
		&       & \multicolumn{10}{c}{$\phi=0.75$}                                              & DM \\
		${\cal S}_{T}^{0}(\lambda_{1}^{0}=1,\lambda_{2}^{0})$ & T=250 & 0.080 & 0.086 & 0.086 & 0.084 & 0.086 & 0.084 & 0.077 & 0.080 & 0.080 & 0.072 & 0.006 \\
		& T=500 & 0.088 & 0.086 & 0.089 & 0.087 & 0.092 & 0.090 & 0.093 & 0.093 & 0.088 & 0.083 & 0.006 \\
		& T=1000 & 0.098 & 0.098 & 0.099 & 0.089 & 0.088 & 0.089 & 0.090 & 0.093 & 0.096 & 0.086 & 0.007 \\
		&       &       &       &       &       &       &       &       &       &       &       & CW \\
		${\cal S}_{T,adj}^{0}(\lambda_{1}^{0}=1,\lambda_{2}^{0})$ & T=250 & 0.099 & 0.106 & 0.106 & 0.107 & 0.110 & 0.110 & 0.109 & 0.116 & 0.120 & 0.126 & 0.055 \\
		& T=500 & 0.102 & 0.100 & 0.104 & 0.104 & 0.108 & 0.108 & 0.110 & 0.114 & 0.117 & 0.124 & 0.055 \\
		& T=1000 & 0.106 & 0.107 & 0.108 & 0.102 & 0.101 & 0.103 & 0.104 & 0.109 & 0.117 & 0.111 & 0.055 \\ \hline
		&       & \multicolumn{10}{c}{$\phi=0.95$}                                              & DM \\
		${\cal S}_{T}^{0}(\lambda_{1}^{0}=1,\lambda_{2}^{0})$ & T=250 & 0.076 & 0.077 & 0.077 & 0.074 & 0.074 & 0.072 & 0.069 & 0.069 & 0.069 & 0.062 & 0.006 \\
		& T=500 & 0.083 & 0.084 & 0.084 & 0.083 & 0.086 & 0.085 & 0.081 & 0.078 & 0.075 & 0.075 & 0.008 \\
		& T=1000 & 0.088 & 0.089 & 0.085 & 0.088 & 0.089 & 0.086 & 0.090 & 0.087 & 0.089 & 0.084 & 0.007 \\
		&       & \multicolumn{10}{c}{}                                                         & CW \\
		${\cal S}_{T,adj}^{0}(\lambda_{1}^{0}=1,\lambda_{2}^{0})$ & T=250 & 0.102 & 0.107 & 0.107 & 0.109 & 0.108 & 0.115 & 0.112 & 0.118 & 0.124 & 0.140 & 0.064 \\
		& T=500 & 0.098 & 0.102 & 0.103 & 0.104 & 0.108 & 0.108 & 0.108 & 0.108 & 0.110 & 0.124 & 0.059 \\
		& T=1000 & 0.098 & 0.099 & 0.096 & 0.098 & 0.101 & 0.101 & 0.106 & 0.105 & 0.112 & 0.114 & 0.056 \\ \hline
		&       & \multicolumn{10}{c}{$\phi=0.98$}                                              & DM \\ 
		${\cal S}_{T}^{0}(\lambda_{1}^{0}=1,\lambda_{2}^{0})$ & T=250 & 0.078 & 0.079 & 0.078 & 0.077 & 0.075 & 0.073 & 0.073 & 0.071 & 0.074 & 0.068 & 0.011 \\
		& T=500 & 0.082 & 0.081 & 0.081 & 0.081 & 0.080 & 0.078 & 0.080 & 0.075 & 0.072 & 0.062 & 0.010 \\
		& T=1000 & 0.085 & 0.086 & 0.090 & 0.088 & 0.087 & 0.090 & 0.088 & 0.086 & 0.084 & 0.073 & 0.007 \\
		&       &       &       &       &       &       &       &       &       &       &       & CW \\
		${\cal S}_{T,adj}^{0}(\lambda_{1}^{0}=1,\lambda_{2}^{0})$ & T=250 & 0.115 & 0.117 & 0.119 & 0.122 & 0.121 & 0.128 & 0.128 & 0.133 & 0.150 & 0.170 & 0.084 \\
		& T=500 & 0.104 & 0.103 & 0.106 & 0.106 & 0.107 & 0.105 & 0.113 & 0.119 & 0.122 & 0.129 & 0.072 \\
		& T=1000 & 0.098 & 0.100 & 0.104 & 0.103 & 0.102 & 0.104 & 0.108 & 0.110 & 0.113 & 0.109 & 0.060 \\
	\end{tabular}
}
	\label{tab:Tab1}
\end{table}

\newpage
\begin{table}[htbp]
	\centering
	\tabcolsep=0.2cm
	\caption{\footnotesize {\bf DGP1} Empirical size of $\overline{{\cal S}}_{T}(\tau_{0}=0.8;\lambda_{2}^{0})$ and $\overline{{\cal S}}_{T,adj}(\tau_{0}=0.8;\lambda_{2}^{0})$ under conditional homoskedasticity and 10\% nominal size}
	\vspace{0.2cm}
\scalebox{0.78}{\begin{tabular}{llcccccccccc}
		& $\lambda_{2}^{0}$ & 0.500 & 0.600 & 0.700 & 0.750 & 0.800 & 0.850 & 0.900 & 0.950 & 1.000 &    \\ \hline
		&       & \multicolumn{9}{c}{$\phi=0.75$}                                          & DM \\
		$\overline{{\cal S}}_{T}(\tau_{0}=0.8;\lambda_{2}^{0})$  & T=250 & 0.085 & 0.088 & 0.086 & 0.082 & 0.075 & 0.076 & 0.061 & 0.042 & 0.047 & 0.006 \\
		& T=500 & 0.090 & 0.092 & 0.088 & 0.084 & 0.085 & 0.080 & 0.066 & 0.056 & 0.065 & 0.006 \\
		& T=1000 & 0.097 & 0.097 & 0.089 & 0.089 & 0.086 & 0.083 & 0.071 & 0.073 & 0.073 & 0.007 \\
		&       &       &       &       &       &       &       &       &       &       & CW \\
		$\overline{{\cal S}}_{T,adj}(\tau_{0}=0.8;\lambda_{2}^{0})$  & T=250 & 0.104 & 0.111 & 0.117 & 0.120 & 0.124 & 0.151 & 0.154 & 0.121 & 0.110 & 0.055 \\
		& T=500 & 0.103 & 0.110 & 0.108 & 0.107 & 0.119 & 0.130 & 0.139 & 0.116 & 0.106 & 0.055 \\
		& T=1000 & 0.108 & 0.108 & 0.103 & 0.106 & 0.108 & 0.118 & 0.121 & 0.113 & 0.099 & 0.055 \\ \hline
		&       & \multicolumn{9}{c}{$\phi=0.95$}                                          & DM \\
		$\overline{{\cal S}}_{T}(\tau_{0}=0.8;\lambda_{2}^{0})$ & T=250 & 0.078 & 0.077 & 0.078 & 0.072 & 0.068 & 0.067 & 0.053 & 0.038 & 0.043 & 0.006 \\
		& T=500 & 0.083 & 0.085 & 0.087 & 0.085 & 0.079 & 0.072 & 0.055 & 0.046 & 0.054 & 0.008 \\
		& T=1000 & 0.086 & 0.086 & 0.088 & 0.081 & 0.084 & 0.077 & 0.067 & 0.058 & 0.069 & 0.007 \\
		&       &       &       &       &       &       &       &       &       &       & CW \\
		$\overline{{\cal S}}_{T,adj}(\tau_{0}=0.8;\lambda_{2}^{0})$  & T=250 & 0.108 & 0.112 & 0.121 & 0.126 & 0.138 & 0.161 & 0.169 & 0.140 & 0.115 & 0.064 \\
		& T=500 & 0.102 & 0.106 & 0.114 & 0.118 & 0.122 & 0.134 & 0.139 & 0.118 & 0.104 & 0.059 \\
		& T=1000 & 0.097 & 0.097 & 0.104 & 0.102 & 0.108 & 0.116 & 0.126 & 0.104 & 0.098 & 0.056 \\ \hline
		&       & \multicolumn{9}{c}{$\phi=0.98$}                                          & DM \\
		$\overline{{\cal S}}_{T}(\tau_{0}=0.8;\lambda_{2}^{0})$  & T=250 & 0.079 & 0.082 & 0.074 & 0.072 & 0.070 & 0.067 & 0.057 & 0.043 & 0.039 & 0.011 \\
		& T=500 & 0.080 & 0.079 & 0.078 & 0.075 & 0.077 & 0.065 & 0.053 & 0.043 & 0.053 & 0.010 \\
		& T=1000 & 0.083 & 0.088 & 0.088 & 0.082 & 0.084 & 0.074 & 0.061 & 0.059 & 0.067 & 0.007 \\
		&       &       &       &       &       &       &       &       &       &       & CW \\
		$\overline{{\cal S}}_{T,adj}(\tau_{0}=0.8;\lambda_{2}^{0})$  & T=250 & 0.119 & 0.127 & 0.131 & 0.144 & 0.157 & 0.189 & 0.206 & 0.174 & 0.143 & 0.084 \\
		& T=500 & 0.104 & 0.109 & 0.110 & 0.116 & 0.134 & 0.148 & 0.159 & 0.129 & 0.119 & 0.072 \\
		& T=1000 & 0.097 & 0.102 & 0.107 & 0.104 & 0.115 & 0.127 & 0.130 & 0.108 & 0.101 & 0.060 \\
	\end{tabular}}
	\label{tab:Tab2}
\end{table}

\newpage 

\begin{table}[htbp]
	\centering
		\tabcolsep=0.2cm
	\caption{\footnotesize {\bf DGP1} Empirical Power of ${\cal S}_{T}^{0}(\lambda_{1}^{0}=1,\lambda_{2}^{0})$ and ${\cal S}_{T,adj}^{0}(\lambda_{1}^{0}=1,\lambda_{2}^{0})$ under conditional homoskedasticity}
		\vspace{0.2cm}
\scalebox{0.78}{\begin{tabular}{lccccccc}
		$\beta$ & -1.500 & -1.750 & -2.000 & -2.250 & -2.500 & -3.000 & -3.500 \\ \hline
		${\cal S}_{T}^{0}(\lambda_{1}^{0},\lambda_{2}^{0})$ & \multicolumn{7}{c}{$\phi_{1}=0.75$} \\
		$\lambda_{2}^{0}=0.80$ & 0.194 & 0.241 & 0.308 & 0.371 & 0.442 & 0.615 & 0.770 \\
		$\lambda_{2}^{0}=0.85$ & 0.213 & 0.272 & 0.349 & 0.427 & 0.519 & 0.685 & 0.831 \\
		$\lambda_{2}^{0}=0.90$ & 0.245 & 0.328 & 0.412 & 0.512 & 0.605 & 0.779 & 0.895 \\
		$\lambda_{2}^{0}=0.95$ & 0.317 & 0.425 & 0.542 & 0.653 & 0.741 & 0.877 & 0.956 \\
		DM    & 0.312 & 0.440 & 0.566 & 0.666 & 0.752 & 0.877 & 0.946 \\
		${\cal S}_{T,adj}^{0}(\lambda_{1}^{0},\lambda_{2}^{0})$ &       &       &       &       &       &       &  \\
		$\lambda_{2}^{0}=0.80$ & 0.350 & 0.447 & 0.565 & 0.671 & 0.758 & 0.895 & 0.966 \\
		$\lambda_{2}^{0}=0.85$ & 0.399 & 0.511 & 0.635 & 0.732 & 0.815 & 0.933 & 0.983 \\
		$\lambda_{2}^{0}=0.90$ & 0.468 & 0.590 & 0.712 & 0.812 & 0.879 & 0.960 & 0.992 \\
		$\lambda_{2}^{0}=0.95$ & 0.579 & 0.708 & 0.819 & 0.897 & 0.941 & 0.986 & 0.998 \\
		CW    & 0.740 & 0.853 & 0.924 & 0.965 & 0.984 & 0.997 & 1.000 \\ \hline
		${\cal S}_{T}^{0}(\lambda_{1}^{0},\lambda_{2}^{0})$ & \multicolumn{7}{c}{$\phi_{1}=0.95$} \\
		$\lambda_{2}^{0}=0.80$ & 0.615 & 0.747 & 0.843 & 0.917 & 0.952 & 0.990 & 0.997 \\
		$\lambda_{2}^{0}=0.85$ & 0.681 & 0.801 & 0.885 & 0.945 & 0.968 & 0.994 & 0.999 \\
		$\lambda_{2}^{0}=0.90$ & 0.753 & 0.860 & 0.924 & 0.967 & 0.984 & 0.998 & 0.999 \\
		$\lambda_{2}^{0}=0.95$ & 0.845 & 0.924 & 0.962 & 0.984 & 0.993 & 0.999 & 1.000 \\
		DM    & 0.857 & 0.925 & 0.959 & 0.979 & 0.989 & 0.998 & 0.999 \\
		${\cal S}_{T,adj}^{0}(\lambda_{1}^{0},\lambda_{2}^{0})$ &       &       &       &       &       &       &  \\
		$\lambda_{2}^{0}=0.80$ & 0.850 & 0.925 & 0.965 & 0.988 & 0.995 & 1.000 & 1.000 \\
		$\lambda_{2}^{0}=0.85$ & 0.885 & 0.949 & 0.980 & 0.992 & 0.997 & 1.000 & 1.000 \\
		$\lambda_{2}^{0}=0.90$ & 0.919 & 0.966 & 0.986 & 0.996 & 0.999 & 1.000 & 1.000 \\
		$\lambda_{2}^{0}=0.95$ & 0.955 & 0.986 & 0.993 & 0.999 & 1.000 & 1.000 & 1.000 \\
		CW    & 0.985 & 0.996 & 0.998 & 1.000 & 1.000 & 1.000 & 1.000 \\ \hline
		${\cal S}_{T}^{0}(\lambda_{1}^{0},\lambda_{2}^{0})$ & \multicolumn{7}{c}{$\phi_{1}=0.98$} \\
		$\lambda_{2}^{0}=0.80$ & 0.853 & 0.927 & 0.964 & 0.984 & 0.992 & 0.999 & 1.000 \\
		$\lambda_{2}^{0}=0.85$ & 0.886 & 0.949 & 0.977 & 0.990 & 0.996 & 0.999 & 1.000 \\
		$\lambda_{2}^{0}=0.90$ & 0.921 & 0.966 & 0.986 & 0.994 & 0.998 & 1.000 & 1.000 \\
		$\lambda_{2}^{0}=0.95$ & 0.956 & 0.981 & 0.994 & 0.997 & 0.999 & 1.000 & 1.000 \\
		DM    & 0.956 & 0.982 & 0.992 & 0.997 & 0.998 & 1.000 & 1.000 \\
		${\cal S}_{T,adj}^{0}(\lambda_{1}^{0},\lambda_{2}^{0})$ &       &       &       &       &       &       &  \\
		$\lambda_{2}^{0}=0.80$ & 0.957 & 0.982 & 0.995 & 0.998 & 0.999 & 1.000 & 1.000 \\
		$\lambda_{2}^{0}=0.85$ & 0.968 & 0.988 & 0.997 & 0.999 & 1.000 & 1.000 & 1.000 \\
		$\lambda_{2}^{0}=0.90$ & 0.981 & 0.992 & 0.998 & 0.999 & 1.000 & 1.000 & 1.000 \\
		$\lambda_{2}^{0}=0.95$ & 0.991 & 0.996 & 0.999 & 1.000 & 1.000 & 1.000 & 1.000 \\
		CW    & 0.997 & 0.999 & 1.000 & 1.000 & 1.000 & 1.000 & 1.000 \\
	\end{tabular}}
	\label{tab:Tab3}
\end{table}

\newpage

\begin{table}[htbp]
	\centering
		\tabcolsep=0.2cm
\caption{\footnotesize {\bf DGP1} Empirical Power of $\overline{{\cal S}}_{T}(\tau_{0}=0.8;\lambda_{2}^{0})$ and $\overline{{\cal S}}_{T,adj}(\tau_{0}=0.8;\lambda_{2}^{0})$ under conditional homoskedasticity}
\scalebox{0.74}{\begin{tabular}{lccccccc}
		$\beta$ & -1.500 & -1.750 & -2.000 & -2.250 & -2.500 & -3.000 & -3.500 \\ \hline
		$\overline{{\cal S}}_{T}(\tau_{0}=0.8;\lambda_{2}^{0})$ & \multicolumn{7}{c}{$\phi_{1}=0.75$} \\
		$\lambda_{2}^{0}=0.80$ & 0.272 & 0.345 & 0.450 & 0.548 & 0.635 & 0.800 & 0.911 \\
		$\lambda_{2}^{0}=0.85$ & 0.363 & 0.475 & 0.600 & 0.695 & 0.785 & 0.904 & 0.965 \\
		$\lambda_{2}^{0}=0.90$ & 0.449 & 0.567 & 0.685 & 0.782 & 0.853 & 0.942 & 0.980 \\
		$\lambda_{2}^{0}=0.95$ & 0.383 & 0.495 & 0.615 & 0.723 & 0.803 & 0.911 & 0.967 \\
		$\lambda_{2}^{0}=1$ & 0.305 & 0.397 & 0.498 & 0.602 & 0.695 & 0.840 & 0.926 \\
		DM    & 0.312 & 0.440 & 0.566 & 0.666 & 0.752 & 0.877 & 0.946 \\
		$\overline{{\cal S}}_{T,adj}(\tau_{0}=0.8;\lambda_{2}^{0})$ &       &       &       &       &       &       &  \\
		$\lambda_{2}^{0}=0.80$ & 0.501 & 0.613 & 0.735 & 0.824 & 0.889 & 0.969 & 0.992 \\
		$\lambda_{2}^{0}=0.85$ & 0.632 & 0.747 & 0.849 & 0.914 & 0.955 & 0.990 & 0.998 \\
		$\lambda_{2}^{0}=0.90$ & 0.704 & 0.815 & 0.892 & 0.945 & 0.973 & 0.994 & 0.999 \\
		$\lambda_{2}^{0}=0.95$ & 0.642 & 0.763 & 0.853 & 0.920 & 0.955 & 0.990 & 0.998 \\
		$\lambda_{2}^{0}=1$ & 0.541 & 0.659 & 0.768 & 0.853 & 0.915 & 0.974 & 0.993 \\
		CW    & 0.740 & 0.853 & 0.924 & 0.965 & 0.984 & 0.997 & 1.000 \\ \hline
		$\overline{{\cal S}}_{T}(\tau_{0}=0.8;\lambda_{2}^{0})$ & \multicolumn{7}{c}{$\phi_{1}=0.95$} \\
		$\lambda_{2}^{0}=0.80$ & 0.771 & 0.870 & 0.925 & 0.967 & 0.982 & 0.997 & 0.999 \\
		$\lambda_{2}^{0}=0.85$ & 0.865 & 0.934 & 0.969 & 0.987 & 0.994 & 0.999 & 1.000 \\
		$\lambda_{2}^{0}=0.90$ & 0.904 & 0.956 & 0.976 & 0.992 & 0.996 & 1.000 & 1.000 \\
		$\lambda_{2}^{0}=0.95$ & 0.876 & 0.936 & 0.966 & 0.986 & 0.993 & 0.999 & 1.000 \\
		$\lambda_{2}^{0}=1$ & 0.802 & 0.889 & 0.939 & 0.970 & 0.985 & 0.997 & 0.999 \\
		DM    & 0.857 & 0.925 & 0.959 & 0.979 & 0.989 & 0.998 & 0.999 \\
		$\overline{{\cal S}}_{T,adj}(\tau_{0}=0.8;\lambda_{2}^{0})$ &       &       &       &       &       &       &  \\
		$\lambda_{2}^{0}=0.80$ & 0.922 & 0.970 & 0.989 & 0.996 & 0.999 & 1.000 & 1.000 \\
		$\lambda_{2}^{0}=0.85$ & 0.960 & 0.987 & 0.995 & 0.999 & 1.000 & 1.000 & 1.000 \\
		$\lambda_{2}^{0}=0.90$ & 0.974 & 0.990 & 0.997 & 0.999 & 1.000 & 1.000 & 1.000 \\
		$\lambda_{2}^{0}=0.95$ & 0.964 & 0.989 & 0.994 & 0.999 & 1.000 & 1.000 & 1.000 \\
		$\lambda_{2}^{0}=1$ & 0.941 & 0.975 & 0.989 & 0.997 & 0.999 & 1.000 & 1.000 \\
		CW    & 0.985 & 0.996 & 0.998 & 1.000 & 1.000 & 1.000 & 1.000 \\ \hline
		$\overline{{\cal S}}_{T}(\tau_{0}=0.8;\lambda_{2}^{0})$ & \multicolumn{7}{c}{$\phi_{1}=0.98$} \\
		$\lambda_{2}^{0}=0.80$ & 0.925 & 0.966 & 0.985 & 0.994 & 0.997 & 1.000 & 1.000 \\
		$\lambda_{2}^{0}=0.85$ & 0.960 & 0.983 & 0.994 & 0.998 & 0.999 & 1.000 & 1.000 \\
		$\lambda_{2}^{0}=0.90$ & 0.972 & 0.990 & 0.996 & 0.999 & 0.999 & 1.000 & 1.000 \\
		$\lambda_{2}^{0}=0.95$ & 0.960 & 0.982 & 0.994 & 0.998 & 0.999 & 1.000 & 1.000 \\
		$\lambda_{2}^{0}=1$ & 0.933 & 0.970 & 0.986 & 0.995 & 0.998 & 1.000 & 1.000 \\
		DM    & 0.956 & 0.982 & 0.992 & 0.997 & 0.998 & 1.000 & 1.000 \\
		$\overline{{\cal S}}_{T,adj}(\tau_{0}=0.8;\lambda_{2}^{0})$ &       &       &       &       &       &       &  \\
		$\lambda_{2}^{0}=0.80$ & 0.978 & 0.992 & 0.997 & 1.000 & 1.000 & 1.000 & 1.000 \\
		$\lambda_{2}^{0}=0.85$ & 0.990 & 0.997 & 0.999 & 1.000 & 1.000 & 1.000 & 1.000 \\
		$\lambda_{2}^{0}=0.90$ & 0.994 & 0.998 & 1.000 & 1.000 & 1.000 & 1.000 & 1.000 \\
		$\lambda_{2}^{0}=0.95$ & 0.991 & 0.997 & 0.999 & 1.000 & 1.000 & 1.000 & 1.000 \\
		$\lambda_{2}^{0}=1$ & 0.984 & 0.995 & 0.998 & 1.000 & 1.000 & 1.000 & 1.000 \\
		CW    & 0.997 & 0.999 & 1.000 & 1.000 & 1.000 & 1.000 & 1.000 \\
	\end{tabular}}
	\label{tab:Tab4}
\end{table}

\newpage

\begin{table}[htbp]
		\centering
	\tabcolsep=0.2cm
	\caption{\footnotesize {\bf DGP2} Empirical Size of ${\cal S}_{T}^{0}(\lambda_{1}^{0}=1,\lambda_{2}^{0})$ and ${\cal S}_{T,adj}^{0}(\lambda_{1}^{0}=1,\lambda_{2}^{0})$ under conditional homoskedasticity}
	\vspace{0.2cm}
\scalebox{0.78}{\begin{tabular}{llccccccccccc}
		& $\lambda_{2}^{0}$ & 0.500 & 0.550 & 0.600 & 0.650 & 0.700 & 0.750 & 0.800 & 0.850 & 0.900 & 0.950 &    \\ \hline
		&       &       &       &       &       &       &       &       &       &       &       & DM \\
		${\cal S}_{T}^{0}(\lambda_{1}^{0}=1,\lambda_{2}^{0})$ & T=250 & 0.055 & 0.055 & 0.055 & 0.054 & 0.054 & 0.050 & 0.047 & 0.046 & 0.043 & 0.037 & 0.002 \\
		& T=500 & 0.066 & 0.065 & 0.065 & 0.063 & 0.065 & 0.061 & 0.060 & 0.055 & 0.051 & 0.044 & 0.002 \\
		& T=1000 & 0.078 & 0.079 & 0.078 & 0.073 & 0.073 & 0.071 & 0.073 & 0.070 & 0.064 & 0.053 & 0.001 \\
		&       &       &       &       &       &       &       &       &       &       &       & CW \\
		${\cal S}_{T,adj}^{0}(\lambda_{1}^{0}=1,\lambda_{2}^{0})$ & T=250 & 0.110 & 0.110 & 0.114 & 0.117 & 0.119 & 0.120 & 0.124 & 0.134 & 0.151 & 0.172 & 0.074 \\
		& T=500 & 0.099 & 0.099 & 0.103 & 0.105 & 0.109 & 0.110 & 0.114 & 0.118 & 0.123 & 0.139 & 0.065 \\
		& T=1000 & 0.105 & 0.104 & 0.104 & 0.102 & 0.103 & 0.102 & 0.110 & 0.115 & 0.120 & 0.127 & 0.066 \\
	\end{tabular}}
	\label{tab:Tab5}
\end{table}%

\begin{table}[htbp]
	\centering
\tabcolsep=0.2cm
\caption{\footnotesize {\bf DGP2} Empirical Size of $\overline{{\cal S}}_{T}(\tau_{0}=0.8;\lambda_{2}^{0})$ and $\overline{{\cal S}}_{T,adj}(\tau_{0}=0.8;\lambda_{2}^{0})$ under conditional homoskedasticity}
\vspace{0.2cm}
\scalebox{0.78}{\begin{tabular}{llcccccccccccc}
		& $\lambda_{2}^{0}$ & 0.500 & 0.550 & 0.600 & 0.650 & 0.700 & 0.750 & 0.800 & 0.850 & 0.900 & 0.950 & 1.000 &    \\ \hline
		&       &       &       &       &       &       &       &       &       &       &       &       & DM \\
		$\overline{{\cal S}}_{T}(\tau_{0}=0.8;\lambda_{2}^{0})$  & T=250 & 0.054 & 0.054 & 0.054 & 0.052 & 0.049 & 0.046 & 0.039 & 0.034 & 0.028 & 0.022 & 0.019 & 0.002 \\
		& T=500 & 0.066 & 0.066 & 0.066 & 0.061 & 0.064 & 0.058 & 0.052 & 0.039 & 0.028 & 0.023 & 0.027 & 0.002 \\
		& T=1000 & 0.074 & 0.074 & 0.070 & 0.070 & 0.069 & 0.064 & 0.061 & 0.049 & 0.037 & 0.034 & 0.043 & 0.001 \\
		&       &       &       &       &       &       &       &       &       &       &       &       & CW \\
		$\overline{{\cal S}}_{T,adj}(\tau_{0}=0.8;\lambda_{2}^{0})$  & T=250 & 0.112 & 0.115 & 0.123 & 0.129 & 0.132 & 0.144 & 0.165 & 0.208 & 0.233 & 0.198 & 0.157 & 0.074 \\
		& T=500 & 0.104 & 0.107 & 0.112 & 0.110 & 0.118 & 0.121 & 0.133 & 0.158 & 0.180 & 0.154 & 0.124 & 0.065 \\
		& T=1000 & 0.100 & 0.104 & 0.101 & 0.105 & 0.109 & 0.111 & 0.117 & 0.137 & 0.151 & 0.129 & 0.111 & 0.066 \\
	\end{tabular}}
	\label{tab:Tab6}
\end{table}%

\newpage

\begin{table}[htbp]
	\centering
\tabcolsep=0.2cm
\caption{\footnotesize {\bf DGP2} Empirical Power of ${\cal S}_{T}^{0}(\lambda_{1}^{0}=1,\lambda_{2}^{0})$ and ${\cal S}_{T,adj}^{0}(\lambda_{1}^{0}=1,\lambda_{2}^{0})$ under conditional homoskedasticity}
\vspace{0.2cm}
\scalebox{0.78}{\begin{tabular}{llccccccccccc}
		& $\lambda_{2}^{0}$ & 0.500 & 0.550 & 0.600 & 0.650 & 0.700 & 0.750 & 0.800 & 0.850 & 0.900 & 0.950 &    \\ \hline
		&       &       &       &       &       &       &       &       &       &       &       & DM \\
		${\cal S}_{T}^{0}(\lambda_{1}^{0}=1,\lambda_{2}^{0})$ & T=250 & 0.520 & 0.572 & 0.620 & 0.668 & 0.724 & 0.778 & 0.835 & 0.885 & 0.929 & 0.971 & 0.884 \\
		& T=500 & 0.784 & 0.836 & 0.877 & 0.914 & 0.945 & 0.967 & 0.986 & 0.994 & 0.998 & 1.000 & 0.998 \\
		& T=1000 & 0.960 & 0.976 & 0.987 & 0.995 & 0.998 & 0.999 & 1.000 & 1.000 & 1.000 & 1.000 & 1.000 \\
		&       &       &       &       &       &       &       &       &       &       &       & CW \\
		${\cal S}_{T,adj}^{0}(\lambda_{1}^{0}=1,\lambda_{2}^{0})$ & T=250 & 0.912 & 0.936 & 0.956 & 0.969 & 0.979 & 0.987 & 0.993 & 0.997 & 0.999 & 1.000 & 1.000 \\
		& T=500 & 0.991 & 0.996 & 0.998 & 1.000 & 1.000 & 1.000 & 1.000 & 1.000 & 1.000 & 1.000 & 1.000 \\
		& T=1000 & 1.000 & 1.000 & 1.000 & 1.000 & 1.000 & 1.000 & 1.000 & 1.000 & 1.000 & 1.000 & 1.000 \\
	\end{tabular}}
	\label{tab:Tab7}
\end{table}

\begin{table}[htbp]
	\centering
\tabcolsep=0.2cm
\caption{\footnotesize {\bf DGP2} Empirical Power of $\overline{{\cal S}}_{T}(\tau_{0}=0.8;\lambda_{2}^{0})$ and $\overline{{\cal S}}_{T,adj}(\tau_{0}=0.8;\lambda_{2}^{0})$ under conditional homoskedasticity}
\vspace{0.2cm}
\scalebox{0.78}{\begin{tabular}{llcccccccccccc}
		& $\lambda_{2}^{0}$ & 0.500 & 0.550 & 0.600 & 0.650 & 0.700 & 0.750 & 0.800 & 0.850 & 0.900 & 0.950 & 1.000 &    \\ \hline
		&       &       &       &       &       &       &       &       &       &       &       &       & DM \\
		$\overline{{\cal S}}_{T}(\tau_{0}=0.8;\lambda_{2}^{0})$  & T=250 & 0.557 & 0.619 & 0.677 & 0.739 & 0.805 & 0.871 & 0.933 & 0.970 & 0.982 & 0.975 & 0.957 & 0.884 \\
		& T=500 & 0.821 & 0.874 & 0.918 & 0.951 & 0.974 & 0.991 & 0.998 & 1.000 & 1.000 & 1.000 & 0.999 & 0.998 \\
		& T=1000 & 0.974 & 0.988 & 0.994 & 0.998 & 1.000 & 1.000 & 1.000 & 1.000 & 1.000 & 1.000 & 1.000 & 1.000 \\
		&       &       &       &       &       &       &       &       &       &       &       &       & CW \\
		$\overline{{\cal S}}_{T,adj}(\tau_{0}=0.8;\lambda_{2}^{0})$  & T=250 & 0.930 & 0.953 & 0.970 & 0.982 & 0.991 & 0.995 & 0.998 & 1.000 & 1.000 & 1.000 & 1.000 & 1.000 \\
		& T=500 & 0.995 & 0.998 & 0.999 & 1.000 & 1.000 & 1.000 & 1.000 & 1.000 & 1.000 & 1.000 & 1.000 & 1.000 \\
		& T=1000 & 1.000 & 1.000 & 1.000 & 1.000 & 1.000 & 1.000 & 1.000 & 1.000 & 1.000 & 1.000 & 1.000 & 1.000 	
	\end{tabular}}
	\label{tab:Tab8}
\end{table}

\begin{table}[htbp]
	\centering
	\tabcolsep=0.2cm
	\caption{Exchange Rate Predictability}
\vspace{0.2cm}
\scalebox{0.78}{	\begin{tabular}{lcccccc}
		& EURUSD & YENUSD & GBPUSD & CHFUSD & AUDUSD & CADUSD \\ \hline
		${\cal S}_{T,adj}^{0}(\lambda_{1}^{0}=1,\lambda_{2}^{0})_{nw}$ &       &       &       &       &       &  \\
		$\lambda_{2}^{0}=0.80$ & 1.000 & 1.000 & 0.360 & 0.924 & 0.764 & 0.997 \\
		$\lambda_{2}^{0}=0.85$ & 1.000 & 0.998 & 0.219 & 0.874 & 0.130 & 0.862 \\
		$\lambda_{2}^{0}=0.90$ & 1.000 & 1.000 & 0.731 & 0.838 & 0.846 & 0.989 \\
		$\lambda_{2}^{0}=0.95$ & 0.997 & 0.997 & 0.814 & 0.755 & 0.711 & 0.945 \\ \hline
		$\overline{{\cal S}}_{T,adj}(\tau_{0};\lambda_{2}^{0})_{nw}$ & \multicolumn{6}{c}{}  \\
		$\tau_{0}=0.80$ & 0.640 & 0.928 & 0.957 & 0.508 & 0.993 & 0.954 \\
		$\tau_{0}=0.90$ & 0.471 & 0.607 & 0.695 & 0.500 & 0.216 & 0.374 \\ \hline
		$DM_{nw}$ & 0.091 & 0.438 & 0.530 & 0.643 & 0.276 & 0.096 \\
		$CW_{nw}$ & 0.048 & 0.248 & 0.299 & 0.532 & 0.155 & 0.044 \\
	\end{tabular}}
	\label{tab:Tab9}
\end{table}

\newpage 

\begin{center}
	{\bf APPENDIX B: PROOFS}
\end{center}

{\small 
	\noindent
	{\bf PROOF OF PROPOSITION 1}. We consider the asymptotic behavior of $Z_{T}(\ell_{1},\ell_{2})$ in (\ref{eq:6}). Rescaling the time axis we write 
	$Z_{T}(\lambda_{1},\lambda_{2})\equiv Z_{T}([(T-k_{0})\lambda_{1}], [(T-k_{0})\lambda_{2}])$ and focus on $Z_{T}(\lambda_{1},\lambda_{2})$. Using $\hat{e}_{j,t+1}^{2}=u_{t+1}^{2}+(\hat{e}_{j,t+1}^{2}-u_{t+1}^{2})$ ($j=1,2$) in (\ref{eq:6}) yields
	\begin{ceqn}
		\begin{align}
			Z_{T}(\lambda_{1},\lambda_{2}) & =  \frac{T-k_{0}}{[(T-k_{0})\lambda_{1}]} 
			\left(
			\frac{\sum_{t=k_{0}}^{k_{0}-1+[(T-k_{0})\lambda_{1}]} u_{t+1}^{2}}{\sqrt{T-k_{0}}}-\frac{[(T-k_{0})\lambda_{1}]}{[(T-k_{0})\lambda_{2}]}\frac{\sum_{t=k_{0}}^{k_{0}-1+[(T-k_{0})\lambda_{2}]} u_{t+1}^{2}}{\sqrt{T-k_{0}}}			
			\right) \nonumber \\
			& +  
			\dfrac{T-k_{0}}{[(T-k_{0})\lambda_{1}]}
			\dfrac{\sum_{t=k_{0}}^{k_{0}-1+[(T-k_{0})\lambda_{1}]}(\hat{e}_{1,t+1}^{2}-u_{t+1}^{2})}{\sqrt{T-k_{0}}} \nonumber \\
			& - \dfrac{T-k_{0}}{[(T-k_{0})\lambda_{2}]}
			\dfrac{\sum_{t=k_{0}}^{k_{0}-1+[(T-k_{0})\lambda_{2}]}(\hat{e}_{2,t+1}^{2}-u_{t+1}^{2})}{\sqrt{T-k_{0}}} \nonumber \\
			& \equiv \dfrac{T-k_{0}}{[(T-k_{0})\lambda_{1}]} \ {\cal N}_{1T}(\lambda_{1},\lambda_{2})+\dfrac{T-k_{0}}{[(T-k_{0})\lambda_{1}]} \ {\cal N}_{2T}(\lambda_{1})-\dfrac{T-k_{0}}{[(T-k_{0})\lambda_{2}]} \ {\cal N}_{3T}(\lambda_{2}).
				\label{eq:38}
\end{align}
\end{ceqn}

\noindent From Assumption A(i) we have $\sup_{\lambda_{1}}|{\cal N}_{2T}(\lambda_{1})|=o_{p}(1)$ and $\sup_{\lambda_{2}}|{\cal N}_{3T}(\lambda_{2})|=o_{p}(1)$. Combining with  
\begin{ceqn}
\begin{align}
\sup_{\lambda_{1},\lambda_{2}}\left|\dfrac{[(T-k_{0})\lambda_{1}]}{[(T-k_{0})\lambda_{2}]}-\dfrac{\lambda_{1}}{\lambda_{2}}\right| & = O(1/(T-k_{0}))	\label{eq:39}
\end{align}
\end{ceqn}
\noindent and 
\begin{ceqn}
	\begin{align}
		\sup_{\lambda_{j}}\left|\dfrac{(T-k_{0})}{[(T-k_{0})\lambda_{j}]}-\dfrac{1}{\lambda_{j}}\right| & = O(1/(T-k_{0})) \ \ \ j=1,2	\label{eq:40}
	\end{align}
\end{ceqn}
\noindent gives
	\begin{ceqn}
	\begin{align}
		Z_{T}(\lambda_{1},\lambda_{2}) & =  \dfrac{1}{\lambda_{1}} 
		\left(
		\frac{\sum_{t=k_{0}}^{k_{0}-1+[(T-k_{0})\lambda_{1}]} u_{t+1}^{2}}{\sqrt{T-k_{0}}}-\dfrac{\lambda_{1}}{\lambda_{2}} \dfrac{\sum_{t=k_{0}}^{k_{0}-1+[(T-k_{0})\lambda_{2}]} u_{t+1}^{2}}{\sqrt{T-k_{0}}}			
		\right) + o_{p}(1). 	\label{eq:41}
	\end{align}
\end{ceqn}
\noindent It is now convenient to reformulate (\ref{eq:41}) as 
	\begin{ceqn}
	\begin{align}
		Z_{T}(\lambda_{1},\lambda_{2}) & =  \dfrac{1}{\lambda_{1}} 
		\left(
		\frac{\sum_{t=k_{0}}^{k_{0}-1+[(T-k_{0})\lambda_{1}]} (u_{t+1}^{2}-\sigma^{2}_{u})}{\sqrt{T-k_{0}}}-\dfrac{\lambda_{1}}{\lambda_{2}} \dfrac{\sum_{t=k_{0}}^{k_{0}-1+[(T-k_{0})\lambda_{2}]} (u_{t+1}^{2}-\sigma^{2}_{u})}{\sqrt{T-k_{0}}}			
		\right) \nonumber \\
		& + \sigma^{2}_{u}\sqrt{T-k_{0}}\left(\dfrac{[(T-k_{0})\lambda_{1}]}{(T-k_{0})\lambda_{1}}
		-\dfrac{[(T-k_{0})\lambda_{2}]}{(T-k_{0})\lambda_{2}}
		\right)+o_{p}(1),
\label{eq:42}
	\end{align}
\end{ceqn}
and note that the second component in the right hand side of (\ref{eq:42}) is $O(1/\sqrt{T-k_{0}})$. We now recall that our setting operates under 
fixed and given magnitudes of $(\lambda_{1},\lambda_{2})$, say $(\lambda_{1}^{0},\lambda_{2}^{0})$ chosen such that $(\lambda_{1}^{0},\lambda_{2}^{0}) \in \Lambda^{0}$.
We have
	\begin{ceqn}
	\begin{align}
		Z_{T}(\lambda_{1}^{0},\lambda_{2}^{0}) & =  \dfrac{1}{\lambda_{1}^{0}} 
		\left(
		\frac{\sum_{t=k_{0}}^{k_{0}-1+[(T-k_{0})\lambda_{1}^{0}]} (u_{t+1}^{2}-\sigma^{2}_{u})}{\sqrt{T-k_{0}}}-\dfrac{\lambda_{1}^{0}}{\lambda_{2}^{0}} \dfrac{\sum_{t=k_{0}}^{k_{0}-1+[(T-k_{0})\lambda_{2}^{0}]} (u_{t+1}^{2}-\sigma^{2}_{u})}{\sqrt{T-k_{0}}}			
		\right) + o_{p}(1).
		\label{eq:43}
	\end{align}
\end{ceqn}
\noindent It follows from Assumptions A(ii)-A(iii), the continuous mapping theorem and Slutsky's theorem that 
\begin{ceqn}
\begin{align}
	Z_{T}^{0}(\lambda_{1}^{0},\lambda_{2}^{0}) \equiv \dfrac{Z_{T}(\lambda_{1}^{0},\lambda_{2}^{0})}{\hat{\sigma}} &  \stackrel{\cal D}{\rightarrow}  \dfrac{1}{\lambda_{1}^{0}}\left(W_{\eta}(\lambda_{1}^{0})-\dfrac{\lambda_{1}^{0}}{\lambda_{2}^{0}} \ W_{\eta}(\lambda_{2}^{0})\right). \label{eq:44}
\end{align}
 \end{ceqn}
\noindent The right hand side of (\ref{eq:44}) is a centered Gaussian random variable with variance 
$v^{0}(\lambda_{1}^{0},\lambda_{2}^{0})=|\lambda_{1}^{0}-\lambda_{2}^{0}|/\lambda_{1}^{0}\lambda_{2}^{0}$ as stated in (\ref{eq:11}). Specifically, the statement in (\ref{eq:44}) is equivalent to $Z_{T}^{0}(\lambda_{1}^{0},\lambda_{2}^{0}) \stackrel{\cal D}{\rightarrow} N(0,|\lambda_{1}^{0}-\lambda_{2}^{0}|/\lambda_{1}^{0}\lambda_{2}^{0})$. This also establishes that ${\cal S}_{T}^{0}(\lambda_{1}^{0},\lambda_{2}^{0})\equiv Z_{T}^{0}(\lambda_{1}^{0},\lambda_{2}^{0})/\sqrt{v^{0}(\lambda_{1}^{0},\lambda_{2}^{0})}\stackrel{\cal D}{\rightarrow} N(0,1)$. \qed \\

\noindent 
{\bf PROOF OF PROPOSITION 2}. We view $Z_{T}(\lambda_{1},\lambda_{2})$ in (\ref{eq:42}) as a functional of $\lambda_{1}$ whose range is determined by the choice of $\tau_{0}$, and for a given $\lambda_{2}=\lambda_{2}^{0}$, satisfying $(\tau_{0},\lambda_{2}^{0})\in \overline{\Lambda}^{0}$.  Assumption A(ii) combined with standard continuous mapping arguments applied to (\ref{eq:42}) yields
\begin{ceqn}
	\begin{align}
Z_{T}(\lambda_{1};\lambda_{2}^{0}) & \stackrel{\cal D}\rightarrow 
\sigma \dfrac{1}{\lambda_{1}}\left(W_{\eta}(\lambda_{1})-\dfrac{\lambda_{1}}{\lambda_{2}^{0}}W_{\eta}(\lambda_{2}^{0})\right).
		\label{eq:45}
	\end{align}
\end{ceqn}
The asymptotic behavior of $\overline{Z}_{T}(\tau_{0};\lambda_{2}^{0})$ in (\ref{eq:9}) 
now follows by appealing to Assumptions A(i)-(iii), the continuity of the average operation and (\ref{eq:45}). Specifically, 
\begin{ceqn}
	\begin{align}
		\overline{Z}_{T} (\tau_{0};\lambda_{2}^{0}) & \stackrel{\cal D}\rightarrow  \frac{1}{1-\tau_{0}} \int_{\tau_{0}}^{1} \left[\frac{W_{\eta}(\lambda_{1})}{\lambda_{1}}-
		\frac{W_{\eta}(\lambda_{2}^{0})}{\lambda_{2}^{0}}\right] \ d\lambda_{1}. 
		\label{eq:46}
	\end{align}
\end{ceqn}
Note that 
\begin{ceqn}
	\begin{align}
		E\left|\int_{\tau_{0}}^{1}\dfrac{W_{\eta}(\lambda_{1})}{\lambda_{1}}d\lambda_{1} \right| & \leq  
		\int_{\tau_{0}}^{1}E\left|\dfrac{W_{\eta}(\lambda_{1})}{\lambda_{1}}\right|d\lambda_{1}=
		\int_{\tau_{0}}^{1}\dfrac{E|W_{\eta}(1)|}{\sqrt{\lambda_{1}}}d\lambda_{1}<\infty
		\label{eq:47}
	\end{align}
\end{ceqn}
so that (\ref{eq:46}) is well defined almost surely and by construction centered Gaussian. It now suffices to obtain its variance. We have
\begin{ceqn}
	\begin{align}
		\frac{1}{(1-\tau_{0})^{2}}
		Cov\left[\int_{\tau_{0}}^{1}\left(\dfrac{W_{\eta}(s_{1})}{s_{1}}-\dfrac{W_{\eta}(\lambda_{2}^{0})}{\lambda_{2}^{0}}\right)ds_{1},\int_{\tau_{0}}^{1}\left(\dfrac{W_{\eta}(s_{2})}{s_{2}}-\dfrac{W_{\eta}(\lambda_{2}^{0})}{\lambda_{2}^{0}}\right)ds_{2}
		\right] 
		& =  \nonumber \\
		\frac{1}{(1-\tau_{0})^{2}}
		\int_{\tau_{0}}^{1}\left[
		\int_{\tau_{0}}^{1} 
		Cov\left[\left(\dfrac{W_{\eta}(s_{1})}{s_{1}}-\dfrac{W_{\eta}(\lambda_{2}^{0})}{\lambda_{2}^{0}}\right),\left(\dfrac{W_{\eta}(s_{2})}{s_{2}}-\dfrac{W_{\eta}(\lambda_{2}^{0})}{\lambda_{2}^{0}}\right)
		\right]
		ds_{2}
		\right]
		d s_{1} 
		& =  \nonumber \\
		\frac{1}{(1-\tau_{0})^{2}} \int_{\tau_{0}}^{1}\int_{\tau_{0}}^{1} 
		\left[
		\dfrac{s_{1} \wedge s_{2}}{s_{1} s_{2}}-\dfrac{s_{1}\wedge \lambda_{2}^{0}}{s_{1}\lambda_{2}^{0}}-
		\dfrac{\lambda_{2}^{0}\wedge s_{2}}{s_{2}\lambda_{2}^{0}}+\dfrac{1}{\lambda_{2}^{0}}
		\right] ds_{1} \ ds_{2},
		\label{eq:48}
	\end{align}
\end{ceqn}
\noindent where we appealed to Fubini's Theorem for interchanging expectations with integration in the second row of (\ref{eq:48}). Standard integral calculus now leads to (\ref{eq:13})-(\ref{eq:14}). \qed \\

\noindent The following lemma collects some key results used in the proofs of Proposition 3 and Corollary 1 on the power properties of the proposed tests under stationarity. \\

\noindent 
{\bf LEMMA A1}. Suppose model (\ref{eq:2}) holds with ${\bm \beta}_{2}={\bm \gamma}/T^{1/4}$. Under Assumption B1 and as $T \rightarrow \infty$ we have 
\begin{enumerate}
	
\Item[(i) ] 
	\begin{ceqn}
		\begin{align}
			\sup_{\lambda} 
			\left|
			\dfrac{\sum_{t=k_{0}}^{k_{0}-1+[(T-k_{0})\lambda]} (\hat{e}_{1,t+1}^{2}-u_{t+1}^{2})}{\sqrt{T-k_{0}}}
		-\lambda  \sqrt{1-\pi_{0}} \ {\bm \gamma}'(\bm Q_{22}-\bm Q_{21} \bm Q_{11}^{-1}\bm Q_{12}){\bm \gamma} \right| & = o_{p}(1),
			\label{eq:49} 
		\end{align}
\end{ceqn}

\vspace{0.8cm}

\Item[(ii)]	
\begin{ceqn}
	\begin{align}
		\sup_{\lambda} 	\left|
			\dfrac{\sum_{t=k_{0}}^{k_{0}-1+[(T-k_{0})\lambda]} (\hat{e}_{2,t+1}^{2}-u_{t+1}^{2})}{\sqrt{T-k_{0}}}
			\right| & = o_{p}(1). 
			\label{eq:50}
		\end{align}
	\end{ceqn}

\end{enumerate}

\noindent
{\bf PROOF OF LEMMA A1}. (i) As we operate under model (\ref{eq:2}) with ${\bm \beta}_{2}={\bm \gamma}/T^{1/4}$ we have $\hat{e}_{1,t+1}-u_{t+1}={\bm x'}_{2,t}{\bm \beta}_{2}-{\bm x'}_{1,t} (\hat{\bm \delta}_{1,t}-{\bm \beta}_{1})$ so that the following identity holds

\begin{ceqn}
	\begin{align}
		\dfrac{\sum_{t=k_{0}}^{k_{0}-1+[(T-k_{0})\lambda]} (\hat{e}_{1,t+1}^{2}-u_{t+1}^{2})}{\sqrt{T-k_{0}}} & =  A_{1T}(\lambda)+A_{2T}(\lambda)-2A_{3T}(\lambda)+2A_{4T}(\lambda)-2A_{5T}(\lambda)
		\label{eq:51}
	\end{align}
\end{ceqn}
\noindent where 

\begin{enumerate}
	\item[] $\displaystyle A_{1T}(\lambda)=T^{-\frac{1}{2}} (T-k_{0})^{-\frac{1}{2}} {\bm \gamma}'\sum_{t} \bm x_{2,t}\bm x_{2,t}' \bm \gamma$,
\item[] $\displaystyle A_{2T}(\lambda)=(T-k_{0})^{-\frac{1}{2}} \sum_{t} (\hat{\bm \delta}_{1,t}-\bm \beta_{1})'{\bm x_{1,t}}{\bm x_{1,t}'}(\hat{\bm \delta}_{1,t}-{\bm \beta}_{1})$
\item[] $\displaystyle A_{3T}(\lambda)=T^{-\frac{1}{4}} (T-k_{0})^{-\frac{1}{2}} \sum_{t} (\hat{\bm \delta}_{1,t}-\bm \beta_{1})'{\bm x_{1,t}}{\bm x_{2,t}'} {\bm \gamma}$
\item[] $\displaystyle A_{4T}(\lambda)=T^{-\frac{1}{4}} (T-k_{0})^{-\frac{1}{2}} {\bm \gamma'} \sum_{t} {\bm x_{2,t} u_{t+1}}$
\item[] $\displaystyle A_{5T}(\lambda)=(T-k_{0})^{-\frac{1}{2}} \sum_{t} (\hat{\bm \delta}_{1,t}-{\bm \beta}_{1})'{\bm x}_{1,t}u_{t+1}$
\end{enumerate}
with $t=k_{0},\ldots,k_{0}-1+[(T-k_{0})\lambda]$ in all of the above summations and below, unless otherwise indicated. \\ 

\noindent For $A_{1T}(\lambda)$, we write 
\begin{align}
A_{1T}(\lambda) & = \sqrt{\dfrac{T-k_{0}}{T}} {\bm \gamma}'
\left(\dfrac{\sum_{t} \bm x_{2,t} \bm x_{2,t}'}{T-k_{0}}-\lambda {\bm Q}_{22}\right){\bm \gamma}+
\sqrt{\dfrac{T-k_{0}}{T}} \lambda {\bm \gamma}'{\bm Q}_{22}{\bm \gamma}
\label{eq:52}
\end{align}
\noindent and as $|\sqrt{(T-[T\pi_{0}])/T}-\sqrt{1-\pi_{0}}|=o(1)$ we have 
\begin{align}
\left|A_{1T}(\lambda)-\lambda \ \sqrt{1-\pi_{0}} \ {\bm \gamma}'{\bm Q_{22}}{\bm \gamma}\right| & \leq 
\sqrt{1-\pi_{0}} \
\|{\bm \gamma}\|^{2}\ \left \lVert \dfrac{\sum_{t} \bm x_{2,t}\bm x_{2,t}'}{T-k_{0}}-\lambda \ {\bm Q}_{22} \right \rVert 
\label{eq:53}
\end{align}
so that Assumption B1(i) directly implies 
\begin{align}
\sup_{\lambda}\left|A_{1T}(\lambda)-\lambda \ \sqrt{1-\pi_{0}} \ {\bm \gamma}'{\bm Q_{22}}{\bm \gamma}\right| & = o_{p}(1).
\label{eq:54}
\end{align}

Before focusing on the remainder quantities we consider the limiting behavior of $(\hat{\bm \delta}_{1,t}-{\bm \beta}_{1})$. Setting $t=[Ts]$ we write

\begin{ceqn}
	\begin{align}
		T^{1/4}(\hat{\bm \delta}_{1,[Ts]}-{\bm \beta}_{1}) & =  \left(\dfrac{\sum_{j=1}^{[Ts]}{\bm x}_{1,j-1}{\bm x}_{1,j-1}'}{T}\right)^{-1}\left(\dfrac{\sum_{j=1}^{[Ts]}{\bm x}_{1,j-1}{\bm x}_{2,j-1}'}{T}\right) {\bm \gamma} \nonumber \\
		& +  T^{-1/4} \left(\dfrac{\sum_{j=1}^{[Ts]}{\bm x}_{1,j-1}{\bm x}_{1,j-1}'}{T}\right)^{-1}\left(\dfrac{\sum_{j=1}^{[Ts]}{\bm x}_{1,j-1} u_{j}}{\sqrt{T}}\right).
		\label{eq:55}
	\end{align}
\end{ceqn}

\noindent For the second term in the right hand side of (\ref{eq:55}) we have 

\begin{align}
\sup_{s} \dfrac{1}{T^{1/4}} \left\lVert \left(\dfrac{\sum_{j=1}^{[Ts]}{\bm x}_{1,j-1}{\bm x}_{1,j-1}'}{T}\right)^{-1}\left(\dfrac{\sum_{j=1}^{[Ts]}{\bm x}_{1,j-1} u_{j}}{\sqrt{T}}\right) \right\rVert & = o_{p}(1)
\label{eq:56}
\end{align}
\noindent
due to Assumptions B1(i)-(ii). For the first term in the right hand side of (\ref{eq:55}) we can write

{\footnotesize
\begin{align}
\left\lVert 
\left(\dfrac{\sum_{j=1}^{[Ts]} \bm x_{1,j-1} \bm x_{1,j-1}'}{T}\right)^{-1}
\left(\dfrac{\sum_{j=1}^{[Ts]} \bm x_{1,j-1} \bm x_{2,j-1}'}{T}\right){\bm \gamma}-{\bm Q}_{11}^{-1}{\bm Q}_{12} {\bm \gamma}\right\rVert &  \nonumber \\
\leq \left\lVert 
\left(\dfrac{\sum_{j=1}^{[Ts]} \bm x_{1,j-1} \bm x_{1,j-1}'}{T}\right)^{-1}
-{\bm Q}_{11}^{-1}\right\rVert  
\left( 
\left\lVert 
\dfrac{\sum_{j=1}^{[Ts]} \bm x_{1,j-1} \bm x_{2,j-1}'}{T} {\bm \gamma}-{\bm Q}_{12}{\bm \gamma}
\right\rVert+\left\lVert  {\bm Q}_{12}{\bm \gamma}\right\rVert \right) &  \nonumber \\
+\left\lVert 
\left(\dfrac{\sum_{j=1}^{[Ts]} \bm x_{1,j-1} \bm x_{1,j-1}'}{T}\right)^{-1}
\right\rVert  
\left\lVert  
\dfrac{\sum_{j=1}^{[Ts]} \bm x_{1,j-1} \bm x_{2,j-1}'}{T} {\bm \gamma}-{\bm Q}_{12}{\bm \gamma}
\right\rVert 
\label{eq:57}
\end{align}
}

\noindent 
so that Assumptions B1(i)-(ii) also ensure that 

\begin{align}
	\sup_{s}  \left\lVert \left(\dfrac{\sum_{j=1}^{[Ts]}{\bm x}_{1,j-1}{\bm x}_{1,j-1}'}{T}\right)^{-1}\left(\dfrac{\sum_{j=1}^{[Ts]}{\bm x}_{1,j-1} {\bm x}_{2,j-1}'}{T}\right){\bm \gamma}-{\bm Q}_{11}^{-1}{\bm Q}_{12}{\bm \gamma} \right\rVert & = o_{p}(1).
	\label{eq:58}
\end{align}
\noindent
Combining (\ref{eq:56}) and (\ref{eq:58}) and using the triangle inequality in (\ref{eq:55}) yields

\begin{align}
\sup_{s} \left \lVert T^{1/4}(\hat{\bm \delta}_{1,[Ts]}-{\bm \beta}_{1}) -{\bm Q}_{11}^{-1}{\bm Q}_{12}{\bm \gamma} 
\right \rVert 
& = o_{p}(1).
	\label{eq:59}
\end{align}

We now focus on $A_{2T}(\lambda)$. Using suitable normalisations and appealing to (\ref{eq:59}), we can express $A_{2T}(\lambda)$ as
\begin{align}
A_{2T}(\lambda) & = \sqrt{\dfrac{T-k_{0}}{T}} (T-k_{0})^{-1}\sum_{t} T^{1/4}(\hat{\bm \delta}_{1,t}-\bm \beta_{1})'{\bm x}_{1,t}{\bm x}_{1,t}'T^{1/4}(\hat{\bm \delta}_{1,t}-\bm \beta_{1}) \nonumber \\
& = \sqrt{\dfrac{T-k_{0}}{T}} (T-k_{0})^{-1}\sum_{t} (T^{1/4}(\hat{\bm \delta}_{1,t}-\bm \beta_{1})-{\bm Q}_{11}^{-1}{\bm Q}_{12}{\bm \gamma})'{\bm x}_{1,t}{\bm x}_{1,t}'T^{1/4}(\hat{\bm \delta}_{1,t}-\bm \beta_{1})
\nonumber \\
& + \sqrt{\dfrac{T-k_{0}}{T}} (T-k_{0})^{-1}  {\bm \gamma}'{\bm Q}_{21}{\bm Q}_{11}^{-1}   \sum_{t} {\bm x}_{1,t}{\bm x}_{1,t}'(T^{1/4}(\hat{\bm \delta}_{1,t}-\bm \beta_{1})-{\bm Q}_{11}^{-1}{\bm Q}_{12}{\bm \gamma}) \nonumber \\
& + \sqrt{\dfrac{T-k_{0}}{T}}  {\bm \gamma}'{\bm Q}_{21}{\bm Q}_{11}^{-1}   \left(\dfrac{\sum_{t} {\bm x}_{1,t}{\bm x}_{1,t}'}{T-k_{0}}-\lambda {\bm Q}_{11}\right) {\bm Q}_{11}^{-1}{\bm Q}_{12}{\bm \gamma} \nonumber \\
& + \lambda \sqrt{\dfrac{T-k_{0}}{T}} {\bm \gamma}'{\bm Q}_{21}{\bm Q}_{11}^{-1}{\bm Q}_{12}{\bm \gamma}.
	\label{eq:60}
\end{align}
\noindent Assumption B1(i) combined with the result in (\ref{eq:59}) give
\begin{align}
\sup_{\lambda} \left|A_{2T}(\lambda)-\lambda \sqrt{1-\pi_{0}} {\bm \gamma}'{\bm Q}_{21}{\bm Q}_{11}^{-1}{\bm Q}_{12}{\bm \gamma}\right| & = o_{p}(1).
	\label{eq:61}
\end{align}

For $A_{3T}(\lambda)$ we write 
\begin{align}
A_{3T}(\lambda) & = 
\sqrt{\dfrac{T-k_{0}}{T}}(T-k_{0})^{-1}\sum (T^{1/4}(\hat{\bm \delta}_{1,t}-\bm \beta_{1})-{\bm Q}_{11}^{-1}{\bm Q}_{12}{\bm \gamma})' {\bm x}_{1,t}{\bm x}_{2,t}'{\bm \gamma} \nonumber \\
& + \sqrt{\dfrac{T-k_{0}}{T}} {\bm \gamma}'{\bm Q}_{21} {\bm Q}_{11}^{-1}
\left(\dfrac{\sum {\bm x}_{1,t}{\bm x}_{2,t}'}{T-k_{0}}-\lambda {\bm Q}_{12} \right){\bm \gamma}
+ \sqrt{\dfrac{T-k_{0}}{T}} \lambda {\bm \gamma}'{\bm Q}_{21} {\bm Q}_{11}^{-1} {\bm Q}_{12} {\bm \gamma}
	\label{eq:62}
\end{align}
so that using (\ref{eq:59}), Assumption B1(i) and the triangle inequality yields 
\begin{align}
\sup_{\lambda}\left|A_{3T}(\lambda)-\lambda \sqrt{1-\pi_{0}} {\bm \gamma}'{\bm Q}_{21}
{\bm Q}_{11}^{-1} {\bm Q}_{12} {\bm \gamma}\right| & = o_{p}(1).
	\label{eq:63}
\end{align}

Next, as an immediate consequence of Assumption B1(ii) we have 
\begin{align}
\sup_{\lambda} |A_{4T}(\lambda)| & =o_{p}(1).
	\label{eq:64}	
\end{align}
Finally, using (\ref{eq:59}) together with Assumption B1(ii) yields
\begin{align}
	\sup_{\lambda} |A_{5T}(\lambda)| & =o_{p}(1).	
	\label{eq:65}
\end{align}

Combining (\ref{eq:54}), (\ref{eq:61}) and (\ref{eq:63})-(\ref{eq:65}) with successive uses of the triangle inequality yields the stated result in Lemma A1(i). The statement in (\ref{eq:50}) follows an identical line of argument as above and details are therefore omitted from the exposition here. \\

\noindent
{\bf PROOF OF PROPOSITION 3}. (i) We initially consider the case of a fixed and non-zero ${\bm \beta}_{2}$ and establish that ${\cal S}^{0}_{T}(\lambda_{1}^{0},\lambda_{2}^{0})\stackrel{p}\rightarrow \infty$. Using (\ref{eq:38}) and appealing to Lemma A1(ii) we have 
\begin{align}
	\dfrac{Z_{T}(\lambda_{1}^{0},\lambda_{2}^{0})}{\sqrt{T-k_{0}}} & =  \dfrac{1}{\lambda_{1}^{0}} \ \dfrac{{\cal N}_{1T}(\lambda_{1}^{0},\lambda_{2}^{0})}{\sqrt{T-k_{0}}}+\dfrac{1}{\lambda_{1}^{0}}\  \dfrac{{\cal N}_{2T}(\lambda_{1}^{0})}{\sqrt{T-k_{0}}} +o_{p}(1).
	\label{eq:66}
\end{align}
\noindent We can now note from (\ref{eq:44})-(\ref{eq:45}) that the first term in the right hand side of (\ref{eq:66}) is $O_{p}(T^{-1/2})$ so that 
\begin{ceqn}
	\begin{align}
		\dfrac{Z_{T}(\lambda_{1}^{0},\lambda_{2}^{0})}{\sqrt{T-k_{0}}} & =  
		\frac{1}{\lambda_{1}^{0}}
		\left[
		\dfrac{\sum_{t=k_{0}}^{k_{0}-1+[(T-k_{0})\lambda_{1}^{0}]} (\hat{e}_{1,t+1}^{2}-u_{t+1}^{2})}{{T-k_{0}}}\right]+o_{p}(1).
		\label{eq:67}
	\end{align}
\end{ceqn}

\noindent It is now straightforward to adapt the result in Lemma A1(i) to a fixed ${\bm \beta}_{2}$ setting and infer that
\begin{ceqn}
	\begin{align}
		\dfrac{\sum_{t=k_{0}}^{k_{0}-1+[(T-k_{0})\lambda_{1}^{0}]} (\hat{e}_{1,t+1}^{2}-u_{t+1}^{2})}{{T-k_{0}}}
		& \stackrel{p}\rightarrow \lambda_{1}^{0} \ {\bm \beta_{2}}'(\bm Q_{22}-\bm Q_{12} \bm Q_{11}^{-1}\bm Q_{12}){\bm \beta_{2}},
		\label{eq:68}
	\end{align}
\end{ceqn}
yielding (for fixed ${\bm \beta}_{2}$)
\begin{ceqn}
	\begin{align}
		\dfrac{Z_{T}^{0}(\lambda_{1}^{0},\lambda_{2}^{0})}{\sqrt{T-k_{0}}}  & \stackrel{p}\rightarrow  
		\frac{1}{\sigma} {\bm \beta_{2}}'(\bm Q_{22}-\bm Q_{12} \bm Q_{11}^{-1}\bm Q_{12}){\bm \beta_{2}}
		\label{eq:69}
	\end{align}
\end{ceqn}
where we also made use of Assumption B1(iii) ensuring that $\hat{\sigma}\stackrel{p}\rightarrow \sigma \in (0, \infty)$. It now follows that 
\begin{ceqn}
	\begin{align}
		\dfrac{{\cal S}_{T}^{0}(\lambda_{1}^{0},\lambda_{2}^{0})}{\sqrt{T-k_{0}}}\equiv 
		\frac{1}{\sqrt{T-k_{0}}} \dfrac{Z_{T}^{0}(\lambda_{1}^{0},\lambda_{2}^{0})}{\sqrt{v^{0}(\lambda_{1}^{0},\lambda_{2}^{0})}} & \stackrel{p}\rightarrow  \dfrac{1}{\sqrt{v^{0}(\lambda_{1}^{0},\lambda_{2}^{0})}} 
		\frac{1}{\sigma} {\bm \beta_{2}}'(\bm Q_{22}-\bm Q_{12} \bm Q_{11}^{-1}\bm Q_{12}){\bm \beta_{2}}
		\label{eq:70}
	\end{align}
\end{ceqn}
\noindent 
with $v^{0}(\lambda_{1}^{0},\lambda_{2}^{0})$ given by (\ref{eq:11}), thus leading to ${\cal S}_{T}^{0}(\lambda_{1}^{0},\lambda_{2}^{0}) \stackrel{p}\rightarrow \infty$ as stated. Proceeding similarly for $\overline{Z}_{T}(\tau_{0};\lambda_{2}^{0})$ we have 
\begin{ceqn}
	\begin{align}
		\dfrac{\overline{\cal S}_{T}(\tau_{0};\lambda_{2}^{0})}{\sqrt{T-k_{0}}}  & \stackrel{p}\rightarrow  
		\dfrac{1}{\sqrt{\overline{v}(\tau_{0};\lambda_{2}^{0})}} \dfrac{1}{\sigma}
		{\bm \beta_{2}}'(\bm Q_{22}-\bm Q_{12} \bm Q_{11}^{-1}\bm Q_{12}){\bm \beta_{2}}
		\label{eq:71}
	\end{align}
\end{ceqn}
\noindent
with $\overline{v}(\tau_{0};\lambda_{2}^{0})$ as in (\ref{eq:13})-(\ref{eq:14}), thus also establishing that 
$\overline{\cal S}_{T}(\tau_{0};\lambda_{2}^{0})\stackrel{p}\rightarrow \infty$. \\

\noindent
(ii) We next focus on the local asymptotic behavior of the two test statistics with ${\bm \beta}_{2}$ parameterised as ${\bm \beta}_{2}={\bm \gamma}/T^{1/4}$. Using (\ref{eq:38}) in conjunction with Lemma A1(i)-(ii) we have
\begin{ceqn}
	\begin{align}
		\dfrac{Z_{T}^{0}(\lambda_{1}^{0},\lambda_{2}^{0})}{\sqrt{v^{0}(\lambda_{1}^{0},\lambda_{2}^{0})}} & =  \frac{1}{\hat{\sigma}} \frac{1}{\lambda_{1}^{0} \sqrt{v^{0}(\lambda_{1}^{0},\lambda_{2}^{0})}}
		\left[
		\dfrac{\sum_{t=k_{0}}^{k_{0}-1+[(T-k_{0})\lambda_{1}^{0}]} (\hat{e}_{1,t+1}^{2}-u_{t+1}^{2})}{\sqrt{T-k_{0}}}\right]+ \nonumber \\
		&   \frac{1}{\hat{\sigma}} \frac{1}{\lambda_{1}^{0}\sqrt{v^{0}(\lambda_{1}^{0},\lambda_{2}^{0})}}
		\left[
		\frac{\sum_{t=k_{0}}^{k_{0}-1+[(T-k_{0})\lambda_{1}^{0}]} u_{t+1}^{2}}{\sqrt{T-k_{0}}}-\frac{\lambda_{1}^{0}}{\lambda_{2}^{0}}\frac{\sum_{t=k_{0}}^{k_{0}-1+[(T-k_{0})\lambda_{2}^{0}]} u_{t+1}^{2}}{\sqrt{T-k_{0}}}
		\right] \nonumber \\
		& + o_{p}(1).
		\label{eq:72}
	\end{align}
\end{ceqn}
\noindent It now follows directly from (\ref{eq:49}) in Lemma A1, Assumption B1(iii) and Slutsky's theorem that 
\begin{ceqn}
	\begin{align}
		{\cal S}_{T}^{0}(\lambda_{1}^{0},\lambda_{2}^{0}) \equiv 
		\dfrac{Z_{T}^{0}(\lambda_{1}^{0},\lambda_{2}^{0})}{\sqrt{v^{0}(\lambda_{1}^{0},\lambda_{2}^{0})}} & \stackrel{\cal D}{\rightarrow}  \dfrac{\sqrt{1-\pi_{0}}}{\sigma \sqrt{v^{0}(\lambda_{1}^{0},\lambda_{2}^{0})}} {\bm \gamma}'(\bm Q_{22}-\bm Q_{12} \bm Q_{11}^{-1}\bm Q_{12}){\bm \gamma}+N(0,1)
		\label{eq:73}
	\end{align}
\end{ceqn}
\noindent as required. The result for $\overline{Z}_{T}(\tau_{0};\lambda_{2}^{0})$ follows identical arguments and is therefore omitted. \qed \\

\noindent
{\bf PROOF OF COROLLARY 1}. Follows directly from (\ref{eq:71})-(\ref{eq:73}) in the proof of Proposition 3. \qed \\

\noindent 
{\bf LEMMA A2}.
\begin{enumerate}
	\item[(i)] Under Assumption B2 and as $T \rightarrow \infty$ we have for $\lambda \in [0,1]$
	 
	\begin{ceqn}
		\begin{align}
			\dfrac{1}{(T-k_{0})^{2}} \sum_{t=k_{0}}^{k_{0}+[(T-k_{0})\lambda]} {\bm x}_{t} {\bm x}_{t}' 
			& \stackrel{\cal D}\rightarrow \dfrac{1}{(1-\pi_{0})^{2}}\int_{\pi_{0}}^{\pi_{0}+(1-\pi_{0})\lambda} {\bm J}_{C}{\bm J}_{C}'dr
			\label{eq:74}
		\end{align}
	\end{ceqn}
	\item[(ii)] Under Assumption B2 and as $T \rightarrow \infty$ we have for $\lambda \in [0,1]$
	\begin{ceqn}
		\begin{align}
			\sup_{\lambda} \left\lVert
			\dfrac{1}{T-k_{0}} \sum_{t=k_{0}}^{k_{0}+[(T-k_{0})\lambda]} {\bm x}_{t} u_{t+1} \right\rVert & = O_{p}(1)
			\label{eq:75}
		\end{align}
	\end{ceqn}
	\item[(iii)] Suppose model (\ref{eq:2}) holds with ${\bm \beta}_{2}={\bm \gamma}/T^{3/4}$. Under Assumption B2 and as $T \rightarrow \infty$ we have 
	\begin{ceqn}
		\begin{align}
			T^{3/4}({\bm \delta}_{1,[Ts]}-{\bm \beta}_{1}) & \stackrel{\cal D}\rightarrow  
			\left(\int_{0}^{s}{\bm J}_{1C}{\bm J}_{1C}'dr\right)^{-1}\left(\int_{0}^{s}{\bm J}_{1C}{\bm J}_{2C}'dr\right)
			{\bm \gamma} \equiv {\bm M}(s) \ {\bm \gamma}.
			\label{eq:76}
		\end{align}
	\end{ceqn}
\end{enumerate}

\noindent
{\bf PROOF OF LEMMA A2}. For (\ref{eq:74}) we have
\begin{ceqn}
	\begin{align}
		\frac{1}{(T-k_{0})^{2}}\sum_{t=k_{0}}^{k_{0}+[(T-k_{0})\lambda]} {\bm x}_{t}{\bm x}_{t}' & = 
		\left(\frac{T}{T-k_{0}}\right)^{2} \ \frac{1}{T^{2}} \sum_{t=k_{0}}^{k_{0}+[(T-k_{0})\lambda]} {\bm x}_{t}{\bm x}_{t}' \nonumber \\
		& = \left(\frac{T}{T-k_{0}}\right)^{2} \ \sum_{t=k_{0}}^{k_{0}+[(T-k_{0})\lambda]} \int_{\frac{t-1}{T}}^{\frac{t}{T}} 
		\left(\dfrac{{\bm x}_{[Tr]}}{\sqrt{T}}\right)\left(\dfrac{{\bm x}_{[Tr]}}{\sqrt{T}}\right)' dr \nonumber \\
		& \stackrel{\cal D}\rightarrow \frac{1}{(1-\pi_{0})^{2}} \int_{\pi_{0}}^{\pi_{0}+(1-\pi_{0})\lambda} {\bm J}_{C}{\bm J}_{C}'dr
		\label{eq:77}
	\end{align}
\end{ceqn}
\noindent due to Assumption B2(i). 
For (\ref{eq:75}) we have
\begin{ceqn}
	\begin{align}
		\sup_{\lambda} \left\lVert
		\dfrac{1}{T-k_{0}} \sum_{t=k_{0}}^{k_{0}+[(T-k_{0})\lambda]} {\bm x}_{t} u_{t+1} \right\rVert & \leq 
		\sup_{\lambda} \left| \dfrac{\sum u_{t+1}}{\sqrt{T-k_{0}}}\right| \sup_{\lambda}
		\left\lVert 
		\dfrac{{\bm x}_{[(T-k_{0})\lambda]}}{\sqrt{T-k_{0}}} \right\rVert \nonumber \\
		& = O_{p}(1)
		\label{eq:78}
	\end{align}
\end{ceqn}
\noindent which also follows from Assumption B2(i). For (\ref{eq:76}) we write 
\begin{ceqn}
	\begin{align}
		T^{3/4}({\bm \delta}_{1,[Ts]}-{\bm \beta}_{1}) & =  
		\left(
		\dfrac{\sum_{j=1}^{[Ts]} {\bm x}_{1,j-1}{\bm x}_{1,j-1}'}{T^{2}}
		\right)^{-1}
		\left(
		\dfrac{\sum_{j=1}^{[Ts]} {\bm x}_{1,j-1}{\bm x}_{2,j-1}'}{T^{2}}
		\right){\bm \gamma} \nonumber \\
		& +  
		\frac{1}{T^{1/4}} \left(
		\dfrac{\sum_{j=1}^{[Ts]} {\bm x}_{1,j-1}{\bm x}_{1,j-1}'}{T^{2}}
		\right)^{-1}
		\left(
		\dfrac{\sum_{j=1}^{[Ts]} {\bm x}_{1,j-1}u_{j}}{T}
		\right).  
		\label{eq:79}
	\end{align}
\end{ceqn}
\noindent From (\ref{eq:75}) it also follows that
\begin{ceqn}
	\begin{align}
		T^{3/4}({\bm \delta}_{1,[Ts]}-{\bm \beta}_{1}) 
		& = \left(
		\dfrac{\sum_{j=1}^{[Ts]} {\bm x}_{1,j-1}{\bm x}_{1,j-1}'}{T^{2}}
		\right)^{-1}
		\left(
		\dfrac{\sum_{j=1}^{[Ts]} {\bm x}_{1,j-1}{\bm x}_{2,j-1}'}{T^{2}}
		\right){\bm \gamma}+o_{p}(1)
		\label{eq:80}
	\end{align}
\end{ceqn}
\noindent and the statement in (\ref{eq:76}) follows directly using (\ref{eq:74}) in (\ref{eq:80}) and appealing to the continuous mapping theorem. \\

\noindent 
{\bf LEMMA A3}.
\begin{enumerate}
	\item[(i)] Suppose model (\ref{eq:2}) holds with ${\bm \beta}_{2}={\bm \gamma}/T^{3/4}$. Under Assumptions B2 and as $T \rightarrow \infty$ we have 
	\begin{ceqn}
		\begin{align}
			\dfrac{\sum_{t=k_{0}}^{k_{0}-1+[(T-k_{0})\lambda]} (\hat{e}_{1,t+1}^{2}-u_{t+1}^{2})}{\sqrt{T-k_{0}}}
			& \stackrel{\cal{D}}\rightarrow  \dfrac{1}{\sqrt{1-\pi_{0}}} \ {\bm \gamma}' 
			\left(
			\int_{\pi_{0}}^{\pi_{0}+(1-\pi_{0})\lambda} {{\bm J}_{C}}^{*}(s) {{{\bm J}_{C}}^{*}(s)}'ds
			\right)  
			{\bm \gamma} 
			\label{eq:81} 
		\end{align}
	\end{ceqn}
	\noindent
	where ${\bm J}_{C}^{*}(s)={\bm J}_{2C}(s)-{\bm M}(s){\bm J}_{1C}(s)$ for ${\bm M}(s)=
	(\int_{0}^{s}{\bm J}_{1C}{\bm J}_{1C}'dr)^{-1}(\int_{0}^{s}{\bm J}_{1C}{\bm J}_{2C}'dr)$. \\
	
	\item[(ii)] Suppose model (\ref{eq:2}) holds with ${\bm \beta}_{2}={\bm \gamma}/T^{3/4}$. Under Assumptions B2 and as $T \rightarrow \infty$ we have  
	\begin{ceqn}
		\begin{align}
			\sup_{\lambda \in (0,1]} 
			\left|
			\dfrac{\sum_{t=k_{0}}^{k_{0}-1+[(T-k_{0})\lambda]} (\hat{e}_{2,t+1}^{2}-u_{t+1}^{2})}{\sqrt{T-k_{0}}}
			\right| & \stackrel{p}\rightarrow  0.
			\label{eq:82}
		\end{align}
	\end{ceqn}
\end{enumerate}

\noindent
{\bf PROOF OF LEMMA A3}. We consider (\ref{eq:81}) first.  We operate under ${\bm \beta}_{2}={\bm \gamma}/T^{3/4}$. Recalling that $\hat{e}_{1,t+1}-u_{t+1}={\bm x}_{2,t}'{\bm \beta}_{2}-{\bm x}_{1,t}'(\hat{\bm \delta}_{1,t}-{\bm \beta}_{1})$ 
and using $\lim_{T\rightarrow \infty} ((T-k_{0})/T)^{j} \rightarrow (1-\pi_{0})^{j}$, we write
\begin{ceqn}
	\begin{align}
		& \dfrac{\sum_{t=k_{0}}^{k_{0}-1+[(T-k_{0})\lambda]} (\hat{e}_{1,t+1}^{2}-u_{t+1}^{2})}{\sqrt{T-k_{0}}} \nonumber \\
		& = 
		\frac{(1-\pi_{0})^{3/2}}{(T-k_{0})^{2}} {\bm \gamma}'\left(\sum_{t=k_{0}}^{k_{0}-1+[(T-k_{0})\lambda]} \bm x_{2,t}\bm x_{2,t}'\right) {\bm \gamma}  \nonumber \\
		& + \frac{(1-\pi_{0})^{3/2}}{(T-k_{0})^{2}}\sum_{t=k_{0}}^{k_{0}-1+[(T-k_{0})\lambda]}
		(T^{3/4}(\hat{\bm \delta}_{1,t}-\bm \beta_{1}))'{\bm x}_{1,t}{\bm x}_{1,t}'
		(T^{3/4}(\hat{\bm \delta}_{1,t}-\bm \beta_{1})) \nonumber \\
		& - 2  \frac{(1-\pi_{0})^{3/2}}{(T-k_{0})^{2}} \left(\sum_{t=k_{0}}^{k_{0}-1+[(T-k_{0})\lambda]} (T^{3/4}(\hat{\bm \delta}_{1,t}-\bm \beta_{1}))' \bm x_{1,t}\bm x_{2,t}' \right) {\bm \gamma} \nonumber \\
		& + 2\frac{(1-\pi_{0})^{1/2}}{T^{1/4}} \left(\frac{1}{T-k_{0}} {\bm \gamma}' \sum_{t=k_{0}}^{k_{0}-1+[(T-k_{0})\lambda]} {\bm x}_{2,t}u_{t+1}\right) \nonumber \\
		& -2 \frac{ (1-\pi_{0})^{1/2}}{T^{1/4}} \left(\frac{1}{T-k_{0}} \sum_{t=k_{0}}^{k_{0}-1+[(T-k_{0})\lambda]} (T^{3/4}(\hat{\bm \delta}_{1,t}-\bm \beta_{1}))' {\bm x}_{1,t}u_{t+1}\right)+o(1)
		\label{eq:83}
	\end{align}
\end{ceqn}

\noindent It next follows from (\ref{eq:75})-(\ref{eq:76}) that the last two terms in the right hand side of (\ref{eq:83}) are $O_{p}(T^{-1/4})$ so that we also have
\begin{ceqn}
	\begin{align}
		& \dfrac{\sum_{t=k_{0}}^{k_{0}-1+[(T-k_{0})\lambda]} (\hat{e}_{1,t+1}^{2}-u_{t+1}^{2})}{\sqrt{T-k_{0}}}= \nonumber \\
		&  \frac{(1-\pi_{0})^{3/2}}{(T-k_{0})^{2}} {\bm \gamma}'\left(\sum_{t=k_{0}}^{k_{0}-1+[(T-k_{0})\lambda]} \bm x_{2,t}\bm x_{2,t}'\right) {\bm \gamma}  \nonumber \\
		& + \frac{ (1-\pi_{0})^{3/2}}{(T-k_{0})^{2}}\sum_{t=k_{0}}^{k_{0}-1+[(T-k_{0})\lambda]}
		(T^{3/4}(\hat{\bm \delta}_{1,t}-\bm \beta_{1}))'{\bm x}_{1,t}{\bm x}_{1,t}'
		(T^{3/4}(\hat{\bm \delta}_{1,t}-\bm \beta_{1})) \nonumber \\
		& - 2 \frac{(1-\pi_{0})^{3/2} }{(T-k_{0})^{2}} \left(\sum_{t=k_{0}}^{k_{0}-1+[(T-k_{0})\lambda]} (T^{3/4}(\hat{\bm \delta}_{1,t}-\bm \beta_{1}))' \bm x_{1,t}\bm x_{2,t}' \right) {\bm \gamma} +o_{p}(1).
		\label{eq:84}
	\end{align}
\end{ceqn}

\noindent Using (\ref{eq:74}) and (\ref{eq:76}) from Lemma A2 together with the continuous mapping theorem, (\ref{eq:84})
leads to the required result in (\ref{eq:81}). The result in (\ref{eq:82}) is established following similar arguments and details are omitted. \qed \\

\noindent
{\bf PROOF OF PROPOSITION 4 and COROLLARY 2}. We focus on Part (ii) of the Proposition as the test consistency property stated in Part (i) follows as its direct consequence. From 
(\ref{eq:38}) and Lemma A2, under the local alternative ${\bm \beta}_{2}={\bm \gamma}/T^{3/4}$ we have 
\begin{ceqn}
	\begin{align}
		& \dfrac{Z_{T}^{0}(\lambda_{1}^{0},\lambda_{2}^{0})}{\sqrt{v^{0}(\lambda_{1}^{0},\lambda_{2}^{0})}} \nonumber \\
		& =  \frac{1}{\hat{\sigma}} \frac{1}{\lambda_{1}^{0}\sqrt{v^{0}(\lambda_{1}^{0},\lambda_{2}^{0})}}
		\left[
		\frac{\sum_{t=k_{0}}^{k_{0}-1+[(T-k_{0})\lambda_{1}^{0}]} u_{t+1}^{2}}{\sqrt{T-k_{0}}}-\frac{\lambda_{1}}{\lambda_{2}^{0}}\frac{\sum_{t=k_{0}}^{k_{0}-1+[(T-k_{0})\lambda_{2}^{0}]} u_{t+1}^{2}}{\sqrt{T-k_{0}}}
		\right]+ \nonumber \\
		& \frac{1}{\hat{\sigma}} \frac{1}{\lambda_{1}^{0} \sqrt{v^{0}(\lambda_{1}^{0},\lambda_{2}^{0})}}
		\left[
		\dfrac{\sum_{t=k_{0}}^{k_{0}-1+[(T-k_{0})\lambda_{1}^{0}]} (\hat{e}_{1,t+1}^{2}-u_{t+1}^{2})}{\sqrt{T-k_{0}}}\right]+o_{p}(1) \nonumber \\
		& \stackrel{\cal D}{\rightarrow} N(0,1) \nonumber \\
		& + \frac{1}{\sigma}\frac{1}{\lambda_{1}^{0}\sqrt{v^{0}(\lambda_{1}^{0},\lambda_{2}^{0})}}
		\frac{1}{\sqrt{1-\pi_{0}}}  {\bm \gamma}' 
		\left(
		\int_{\pi_{0}}^{\pi_{0}+(1-\pi_{0})\lambda_{1}^{0}} {{\bm J}_{C}}^{*}(s) {{{\bm J}_{C}}^{*}(s)}'ds
		\right) 
		{\bm \gamma} 
		\label{eq:85} 
	\end{align}
\end{ceqn}
\noindent
using (\ref{eq:81}), Slutsky and the continuous mapping theorems (note that the standard normality of the first component in the right hand side of (\ref{eq:85}) has been established in Proposition 1). It now follows directly from (\ref{eq:85}) that $\lim_{||\gamma||\rightarrow \infty}\lim_{T\rightarrow \infty} {\cal S}_{T}(\lambda_{1}^{0},\lambda_{2}^{0})$ as required. The result for $\overline{\cal S}_{T}(\tau_{0};\lambda_{2}^{0})$ follows identical lines and its details are omitted.  \qed \\

\noindent
{\bf PROOF OF PROPOSITION 5 and COROLLARY 3}. We have $\hat{e}_{1,t+1}^{2}-\tilde{e}_{2,t+1}^{2}=(\hat{e}_{1,t+1}^{2}-
\hat{e}_{2,t+1}^{2})+(\hat{e}_{1,t+1}-\hat{e}_{2,t+1})^{2}$ 
which leads to the formulations of ${\cal S}_{T,adj}^{0}(\lambda_{1}^{0},\lambda_{2}^{0})$
and $\overline{\cal S}_{T,adj}(\tau_{0};\lambda_{2}^{0})$ in (\ref{eq:34}) and (\ref{eq:35}) respectively. Under the null hypothesis and for both test statistics the result follows 
by verifying that 
\begin{ceqn}
	\begin{align}
		\sup_{\lambda} \left|
		\dfrac{ \sum_{t=k_{0}}^{k_{0}-1+[(T-k_{0})\lambda]}(\hat{e}_{1,t+1}-\hat{e}_{2,t+1})^{2}}{\sqrt{T-k_{0}}}
		\right| & \stackrel{p} \rightarrow  0.
		\label{eq:86}
	\end{align}
\end{ceqn}
\noindent Noting that
\begin{ceqn} 
	\begin{align}
		\dfrac{ \sum_{t=k_{0}}^{k_{0}-1+[(T-k_{0})\lambda]}(\hat{e}_{1,t+1}-\hat{e}_{2,t+1})^{2}}{\sqrt{T-k_{0}}} & =  
		\dfrac{\sum_{t=k_{0}}^{k_{0}-1+[(T-k_{0})\lambda]} (\hat{e}_{1,t+1}^{2}-u_{t+1}^{2})}{\sqrt{T-k_{0}}}+ \nonumber \\
		& 
		\dfrac{\sum_{t=k_{0}}^{k_{0}-1+[(T-k_{0})\lambda]} (\hat{e}_{2,t+1}^{2}-u_{t+1}^{2})}{\sqrt{T-k_{0}}} \nonumber \\
		& - 2\dfrac{\sum_{t=k_{0}}^{k_{0}-1+[(T-k_{0})\lambda]} (\hat{e}_{1,t+1}\hat{e}_{2,t+1}-u_{t+1}^{2})}{\sqrt{T-k_{0}}},
		\label{eq:87}
	\end{align}
\end{ceqn}
the statement in (\ref{eq:86}) follows directly from Assumption A(i) since we operate  under the null hypothesis noting also 
that $(\hat{e}_{1,t+1}\hat{e}_{2,t+1}-u_{t+1}^{2})=(\hat{e}_{1,t+1}-\hat{e}_{2,t+1})\hat{e}_{2,t+1}+(\hat{e}_{2,t+1}^{2}-u_{t+1}^{2})$ from which we infer the $o_{p}(1)$'ness of the third component in the right hand side of (\ref{eq:87}).  
It now follows that Propositions 1 and 2 continue to hold for the two adjusted statistics. \\

\noindent
For the behavior of the adjusted statistics under the alternative we initially consider the case of stationary predictors and operate under 
${\bm \beta}_{2}={\bm \gamma}/T^{1/4}$ as in the setting of Corollary 1. Using (\ref{eq:87}) with Lemma A1 we can write 
\begin{ceqn}
	\begin{align}
		h_{T}^{0}(\lambda_{1}^{0},\lambda_{2}^{0}) & = \frac{1}{\hat{\sigma}}\frac{1}{\sqrt{v^{0}(\lambda_{1}^{0},\lambda_{2}^{0})} \lambda_{2}^{0}}
		\dfrac{\sum_{t=k_{0}}^{k_{0}-1+[(T-k_{0})\lambda_{2}^{0}]} (\hat{e}_{1,t+1}^{2}-u_{t+1}^{2})}{\sqrt{T-k_{0}}}+o_{p}(1) \nonumber \\
		& \stackrel{p}\rightarrow 
		\frac{1}{\sigma}\frac{1}{\sqrt{v^{0}(\lambda_{1}^{0},\lambda_{2}^{0})}}\sqrt{1-\pi_{0}}{\bm \gamma}'(\bm Q_{22}-\bm Q_{12} \bm Q_{11}^{-1}\bm Q_{12}){\bm \gamma} \equiv \psi^{0}
		\label{eq:88}
	\end{align}
\end{ceqn}
\noindent
and 
\begin{ceqn}
	\begin{align}
		\overline{h}_{T}(\tau_{0};\lambda_{2}^{0}) & = \frac{1}{\hat{\sigma}}\frac{1}{\sqrt{\overline{v}(\tau_{0};\lambda_{2}^{0})} \lambda_{2}^{0}}
		\dfrac{\sum_{t=k_{0}}^{k_{0}-1+[(T-k_{0})\lambda_{2}^{0}]} (\hat{e}_{1,t+1}^{2}-u_{t+1}^{2})}{\sqrt{T-k_{0}}}+o_{p}(1) \nonumber \\
		& \stackrel{p}\rightarrow 
		\frac{1}{\sigma}\frac{1}{\sqrt{\overline{v}(\tau_{0};\lambda_{2}^{0})}}\sqrt{1-\pi_{0}}{\bm \gamma}'(\bm Q_{22}-\bm Q_{12} \bm Q_{11}^{-1}\bm Q_{12}){\bm \gamma} \equiv \overline{\psi}.
		\label{eq:89}
	\end{align}
\end{ceqn}

\noindent Using (\ref{eq:88})-(\ref{eq:89}) it follows that ${\cal S}_{T,adj}^{0}(\lambda_{1}^{0},\lambda_{2}^{0})\stackrel{\cal D}\rightarrow N(2\psi^{0},1)$ and similarly for 
$\overline{\cal S}_{T,adj}^{0}(\lambda_{1}^{0},\lambda_{2}^{0})\stackrel{\cal D}\rightarrow N(2\overline{\psi},1)$ which establishes the fact that Proposition 3 continues to hold for the two adjusted statistics in addition to part (i) of Corollary 3. 
The result for the case of persistent predictors follows identical lines, making use of Lemma A3 and Proposition 3
which in turn establishes part (ii) of Corollary 3 and that Proposition 4 also holds for the two adjusted statistics.  \qed 
}

\vspace{0.08cm}

\section*{Online Supplementary Material}
\noindent
Pitarakis, Jean-Yves (2023): Supplement to ``A novel approach to predictive Accuracy Testing in Nested Environments'', Econometric Theory Supplementary Material. To view, please visit:

\end{document}